\documentclass[12pt]{article}

\usepackage{amssymb, graphicx, natbib, parskip}

\bibpunct[,]{(}{)}{;}{a}{}{,}

\newcommand{\aap}{A\&A}

\newcommand{\apj}{ApJ}
\newcommand{\apjl}{ApJ}
\newcommand{\apjs}{ApJS}
\newcommand{\araa}{ARA\&A}
\newcommand{\mnras}{MNRAS}
\newcommand{\nat}{Nature}
\newcommand{\nar}{New Astron. Rev.}

\newcommand{\jcap}{JCAP}
\newcommand{\pasj}{PASJ}
\newcommand{\physrep}{Phys. Rep.}
\newcommand{\planss}{Planet. Space Sci.}
\newcommand{\prd}{Phys. Rev. D}

\newcommand{\sovast}{SvA}
\newcommand{\ssr}{Space Sci. Rev.}

\begin{document}

\title{The numerical frontier of the high-redshift Universe}

\author{Thomas H. Greif\thanks{E-Mail: tgreif@cfa.harvard.edu}}

\date{\normalsize Harvard-Smithsonian Center for Astrophysics}

\maketitle

\section*{Abstract}

The first stars are believed to have formed a few hundred million years after the big bang in so-called dark matter minihalos with masses $\sim 10^6\,M_\odot$. Their radiation lit up the Universe for the first time, and the supernova explosions that ended their brief lives enriched the intergalactic medium with the first heavy elements. Influenced by their feedback, the first galaxies assembled in halos with masses $\sim 10^8\,M_\odot$, and hosted the first metal-enriched stellar populations. In this review, I summarize the theoretical progress made in the field of high-redshift star and galaxy formation since the turn of the millennium, with an emphasis on numerical simulations. These have become the method of choice to understand the multi-scale, multi-physics problem posed by structure formation in the early Universe. In the first part of the review, I focus on the formation of the first stars in minihalos - in particular the post-collapse phase, where disk fragmentation, protostellar evolution, and radiative feedback become important. I also discuss the influence of additional physical processes, such as magnetic fields and streaming velocities. In the second part of the review, I summarize the various feedback mechanisms exerted by the first stars, followed by a discussion of the first galaxies and the various physical processes that operate in them.

\newpage

\section{Introduction}

The formation of the first stars marked a fundamental transition in the history of the Universe. They initiated the transformation of the homogeneous intergalactic medium (IGM) to one filled with the rich structure we observe today. Ending the so-called `cosmic dark ages', when the Universe contained no visible light, they lit up the Universe at redshifts $z\gtrsim 20$ \citep{bl04a, glover05, bromm13, glover13}. They formed at the center of dark matter (DM) `minihalos' with virial masses $M_{\rm vir}\sim 10^6\,{\rm M}_\odot$, which are the smallest building blocks in the hierarchy of galaxy formation. These objects accrete the pure hydrogen and helium gas forged in the Big Bang, and after continued cooling and contraction form a stellar embryo that begins to accrete from the surrounding gas cloud.

Since the virial temperature of minihalos is not high enough to activate atomic hydrogen cooling, the gas can only cool via molecular hydrogen (H$_2$). The importance of H$_2$ for the cooling of low-mass gas clumps that condense out of the expanding Universe was recognized in the late 1960's \citep{sz67, pd68, hirasawa69, matsuda69, takeda69}. Later on, simplified one-zone models accounted for the dynamics of collapsing gas clouds next to the radiative cooling of the gas \citep{yoneyama72, hutchins76, silk77, carlberg81, kr83, pss83, silk83, cba84, ik84, cr86, sun96, uehara96, tegmark97, ns99, vs03, vs05a}. The first one-dimensional calculations of the simultaneous collapse of DM and gas were carried out in the context of the cold dark matter (CDM) paradigm \citep{htl96, on98, nu99}, while three-dimensional simulations had to await improvements in numerical simulation techniques in the late 1990`s \citep{abel98, bcl99, bcl02, abn00, abn02}.

One of the main results of these studies is that the minimum temperature to which the gas can cool via H$_2$ lines is more than an order of magnitude higher than in present-day star formation regions, resulting in a greatly increased Jeans mass. Since the accretion rate is directly related to the Jeans mass, Population~III stars are expected to have masses of the order of $M_*\sim 100\,{\rm M}_\odot$. Following the introduction of this `standard model` of primordial star formation, recent work has focused on refining this picture. In particular, the influence of fragmentation, protostellar interactions, radiation, and magnetic fields were initially neglected. These may have a substantial effect on the collapse of the gas. In this review, I summarize the progress made on these topics since the advent of the first three-dimensional simulations of primordial star formation.

Despite the complications brought about by these processes, it is likely that Population~III stars were much more massive than present-day stars. They are therefore strong emitters of ultraviolet (UV) radiation, which heats the IGM and begins the process of reionization \citep{lb01, cf05, fck06, stiavelli09}. Their radiation also dissociates H$_2$ on cosmological scales and may suppress star formation until halos massive enough to activate atomic hydrogen cooling assemble \citep{cf05}. Next to radiative feedback, the supernova (SN) explosions of massive Population~III stars exert mechanical as well as chemical feedback \citep[e.g.][]{wa08b, greif10}. The ejected metals facilitate the transition to Population~I/II star formation by allowing the gas to cool to much lower temperatures than would otherwise be possible \citep[e.g.][]{omukai00, bromm01}.

Due to the intricate nature of feedback, the formation of the first galaxies in so-called `atomic cooling halos' is much more complicated than the formation of the first stars in minihalos \citep{by11, lf13}. Fully self-consistent simulations that predict the properties of the first galaxies are not yet available. Nevertheless, recent work has shown that supersonic turbulence is one of the key factors governing the properties of the first galaxies \citep[e.g.][]{wa07b, greif08}. Radiative and chemical feedback from stars in progenitor halos influence their formation as well \citep[e.g.][]{wa08b, greif10}. The degree of sophistication of first galaxy simulations has continued to increase, with the most recent studies including star formation and feedback recipes that rival those of present-day galaxy formation simulations \citep[e.g.][]{wise12a}.

The organization of this review is as follows. In Section~\ref{sec_sm}, I give a brief introduction to structure formation in the high-redshift Universe, followed by a description of the collapse of gas in minihalos. I then review the influence of fragmentation on the accretion phase, the evolution of the nascent protostellar system, and the effects of protostellar radiation on the initial mass function (IMF) of the first stars (Section~\ref{sec_nsm}). In Section~\ref{sec_ap}, I discuss additional physics that may affect the formation of the first stars, including hydrogen deuteride (HD) cooling, magnetic fields, cosmic rays, streaming velocities, DM annihilation, and alternative cosmologies. Finally, in Section~\ref{sec_fg} I discuss the radiative, mechanical and chemical feedback exerted by Population~III stars, next to the progress made with respect to the formation of the first galaxies. I focus on theoretical work, with a particular emphasis on numerical simulations, but briefly mention neglected processes and empirical signatures in Section~\ref{sec_es}. The summary is presented in Section~\ref{sec_con}. All distances are quoted in proper units, unless noted otherwise.

Finally, a reference to related reviews is in order. A general description of structure formation in the early Universe may be found in \citet{bl01}, \citet{lfe08}, and \citet{wmb13}. The high-redshift IGM is discussed in \citet{bl07} and \citet{meiksin09}, and the effects of the relative velocity offset between DM and baryons may be found in \citet{fialkov14b}. The formation of the first stars in minihalos is reviewed in \citet{bl04a}, \citet{glover05}, \citet{bromm13}, and \citet{glover13}, while the properties of the first galaxies are described in \citet{by11}, \citet{johnson13c}, and \citet{lf13}. Less focused reviews of star formation at high redshifts may be found in \citet{bromm09} and \citet{loeb10}. Finally, feedback by Population~III is summarized in \citet{cf05}.

\section{First stars: the initial collapse} \label{sec_sm}

\subsection{Structure formation}

On the largest scales, the Universe appears nearly uniform and isotropic. However, the presence of stars and galaxies indicates that below a certain scale the Universe must have begun to deviate from its uniformity. It is believed that these structures grew from infinitesimally small perturbations seeded by quantum fluctuations in the very early Universe. Within variants of the CDM model, where the mass density of the Universe is dominated by DM, the matter overdensity $\delta=\left(\rho-{\bar\rho}\right)/{\bar\rho}$, where $\rho$ is the local mass density and ${\bar\rho}$ the mean density of the Universe, grows in proportion to the scale factor $a=1/\left(1+z\right)$, where $z$ denotes redshift. Once the overdensity becomes of order unity, the associated region decouples from the expanding background Universe and collapses under its own gravity. The collapse may occurs simultaneously in one, two, or three dimensions. Structures that collapse in one dimension are termed `sheets', the collapse of two sheets results in a `filament', and DM `halos' form at the intersection of filaments. The likelihood of formation decreases with increasing dimensionality, such that there are many more sheets than halos. However, halos provide the deepest potential wells and are therefore the sites of star and galaxy formation.

\subsection{Dark matter halos}

One of the most important characteristics of the CDM paradigm is the hierarchical, `bottom-up' nature of collapse. Increasingly massive halos form via accretion and merging of low-mass halos. The smallest collapse scale is set by the free-streaming length, which is about $1\,{\rm au}$ if DM consists of a weakly interacting massive particle \citep[WIMP;][]{bvs08}. Following the collapse of a DM halo, it achieves virial equilibrium within a few dynamical times, with the virial density given by $\rho_{\rm vir}=\Delta_{\rm vir}{\bar\rho}$, where $\Delta_{\rm vir}\simeq 200$ \citep[e.g.][]{bl01}. The balance between gravitational potential energy and kinetic energy allows a virial velocity to be defined, and is given by $v_{\rm vir}^2=GM_{\rm vir}/R_{\rm vir}$, where $G$ is the gravitational constant, $M_{\rm vir}$ the virial mass, and $R_{\rm vir}$ the virial radius of the halo. The latter is related to the mass of the halo via $R_{\rm vir}^3=3M_{\rm vir}/\left(4\pi\Delta_{\rm vir}{\bar\rho}\right)$. Halos with virial masses up to $\sim 10^6\,{\rm M}_\odot$ are commonly referred to as minihalos, since they are massive enough to cool and host star formation. A $2$--$3\sigma$ peak typically forms at $z\sim 20$, and their virial radius may be conveniently written as:
\begin{equation}
R_{\rm vir}\simeq 100\,{\rm pc}\left(\frac{M_{\rm vir}}{10^6\,{\rm M}_\odot}\right)^{1/3}\left(\frac{1+z}{20}\right)^{-1}.
\end{equation}
A number of studies have systematically investigated the properties of minihalos, finding that they are usually denser and more clustered than their high-mass counterparts \citep{jh01, heitmann06, lukic07, reed07, cw08, dn09, dn10, ddn11, desouza13a, sasaki14}.

The DM sets the gravitational potential for the gas, which virializes within the halo and heats to the virial temperature $T_{\rm vir}=\mu m_{\rm H}v_{\rm vir}^2/k_{\rm B}$, where $\mu$ is the mean molecular weight, $m_{\rm H}$ the mass of the hydrogen atom, and $k_{\rm B}$ Boltzmann's constant. In terms of a typical minihalo, the virial temperature may be written as:
\begin{equation} \label{eq_tvir}
T_{\rm vir}\simeq 2\times 10^3\,{\rm K}\left(\frac{M_{\rm vir}}{10^6\,{\rm M}_\odot}\right)^{2/3}\left(\frac{1+z}{20}\right).
\end{equation}
In contrast to the pressure-less DM, a minimum scale exists below which the gas cannot collapse. This scale is approximately given by the Jeans length, $\lambda_{\rm J}=c_{\rm s}t_{\rm ff}$, where $c_{\rm s}$ is the sound speed of the gas, and $t_{\rm ff}=\sqrt{3\pi/\left(32G\rho\right)}$ the free-fall time. The corresponding Jeans mass is given by $M_{\rm J}=\pi\rho\lambda_{\rm J}^3/6$, and may be derived from the density and temperature of the IGM. Since the gas temperature is coupled to the cosmic microwave background (CMB) by Compton scattering at $z\gtrsim 100$, the cosmological Jeans mass initially remains constant at $M_{\rm J}\simeq 10^5\,{\rm M}_\odot$ \citep{bl04a}. At lower redshifts, the gas expands adiabatically, and the Jeans mass is given by $M_{\rm J}\simeq 10^4\,{\rm M}_\odot \left[\left(1+z\right)/20\right]^{3/2}$. This is many orders of magnitude larger than the minimum DM halo mass. More careful calculations that use perturbation theory to compute the growth of density fluctuations define a filtering mass below which the gas content in a halo is significantly suppressed with respect to the cosmic mean \citep[e.g.][]{gh98, nb07}. These calculations find that the filtering mass remains nearly constant at a few times $10^4\,{\rm M}_\odot$ at $z\lesssim 100$. Even though this is the minimum mass of a halo at which the gas can contract by a substantial amount, it is not yet a sufficient criterion for star formation.

\subsection{H$_2$ formation and cooling} \label{sec_h2}

For the collapse of the gas to continue beyond the formation of a hydrostatic object, it must be able to radiate away its thermal energy. One of the most efficient coolants in primordial gas is collisional excitation cooling of atomic hydrogen, also termed Ly-$\alpha$ cooling. This coolant is most effective at temperatures $\simeq 10^4\,{\rm K}$, where the first excited state of atomic hydrogen begins to be populated. However, halos with masses greater than the filtering mass, but below approximately $M_{\rm vir}=5\times 10^7\,{\rm M}_\odot$, which corresponds to a virial temperature of $\simeq 10^4\,{\rm K}$ at $z=10$, must rely on another coolant. \citet{sz67} and \citet{pd68} realized that molecular hydrogen (H$_2$) may be such a coolant.

The rotational and vibrational states of H$_2$ in the electronic ground state are excited by collisions with other particles, which decay and allow the gas to cool. The most important transitions operate in the temperature range $200\,{\rm K}\lesssim T\lesssim 5000\,{\rm K}$. At higher temperatures, Ly-$\alpha$ cooling becomes dominant. Despite its importance at high redshifts, the H$_2$ molecule is not a particularly efficient coolant. Since the hydrogen molecule is symmetric, it does not have a permanent electric dipole moment, resulting in correspondingly lower transition probabilities. In addition, the existence of ortho and para states for hydrogen rules out the lowest energy transition $J=1\rightarrow 0$, where $J$ is the rotational quantum number. The least energetic transition with a non-negligible probability of occurring is the $J=2\rightarrow 0$ transition, which corresponds to a temperature of $\simeq 500\,{\rm K}$. In practice, the Maxwellian tail of the velocity distribution allows the gas to cool to $\simeq 200\,{\rm K}$. The absence of a dipole moment in the H$_2$ molecule also rules out its simplest formation channel: the direct association of two hydrogen atoms. The excess energy of the collision cannot be radiated away quickly enough, and so the intermediate system decays again \citep{gs63}.

For this reason, alternative formation channels have been considered \citep{htl96, abel97, gp98, sld98}. For the purpose of primordial star formation, the most important formation channel is via the intermediary reaction of radiative association of free electrons with neutral hydrogen atoms \citep{mcdowell61, pd68}:
\begin{equation}
{\rm H}+{\rm e}^-\rightarrow {\rm H}^-+\gamma.
\end{equation}
Associative detachment of the H$^-$ atoms with neutral hydrogen atoms then results in the formation of H$_2$:
\begin{equation}
{\rm H}^-+{\rm H}\rightarrow {\rm H}_2+{\rm e}^-.
\end{equation}
One of the most important parameters that governs the amount of H$_2$ that can be formed is the electron abundance. It decreases from its relic abundance by recombining with ionized hydrogen:
\begin{equation}
{\rm H}^+ +{\rm e}^-\rightarrow{\rm H}+\gamma.
\end{equation}
Most of the H$_2$ therefore forms within a recombination time. A straightforward calculation shows that the asymptotic H$_2$ abundance depends primarily on the ratio of the H$^-$ formation to the recombination rate constant \citep{glover13}. If the recombination rate is significantly smaller than the radiative association rate, more electrons remain to form H$^-$. For typical values of the rate equations, the asymptotic H$_2$ abundance scales approximately as $y_{\rm H_2}\propto T^{1.5}$, and for $T=1000\,{\rm K}$ lies in the range of $y_{\rm H_2}\simeq 10^{-4}-10^{-3}$ \citep{tegmark97}. This leads to the somewhat counterintuitive result that a more massive halo with a higher virial temperature forms more H$_2$, which allows the gas to cool more efficiently.

The H$_2$ abundance necessary to cool the gas and facilitate its collapse to high densities may be obtained by comparing the cooling time with the Hubble time. The latter estimates the time scale on which the virial temperature of a halo changes significantly \citep{ro77, silk77}. A straightforward calculation shows that for a virial temperature of about $1000\,{\rm K}$, enough H$_2$ is produced to cool the gas \citep{glover13}. This value does not depend strongly on redshift, such that the minimum halo mass required for efficient cooling may be obtained from equation~\ref{eq_tvir}:
\begin{equation}
M_{\rm cool}\simeq 3.5\times 10^5\,{\rm M}_\odot\left(\frac{1+z}{20}\right)^{-3/2}.
\end{equation}
This is typically an order of magnitude higher the filtering mass, showing that not all halos that are massive enough to allow the gas to collapse are also massive enough to facilitate cooling and star formation.

It is important to note that the above derivation of the cooling mass is only approximately valid, since in reality the gas is constantly heated by minor mergers \citep{yoshida03b}. Furthermore, even though the first stars typically formed at $z\sim 20$, they are not the very first stars in the observable Universe. Instead, these are believed to have formed at $z\simeq 50-65$ \citep{gao05, nnb06}.

\subsection{Jeans instability}

During the initial stages of the collapse, the level populations of H$_2$ are not yet populated according to local thermal equilibrium (LTE), and the cooling rate scales as $\Lambda_{\rm H_2}\propto n_{\rm H}^2$, where $n_{\rm H}$ is the number density of hydrogen nuclei. Once enough H$_2$ has formed, a runaway process ensues in which cooling allows the central gas cloud to contract, which results in even more cooling. This collapse phase ends when collisions between H$_2$ molecules and other species become so frequent that the ro-vibrational levels of H$_2$ are populated according to LTE. In this case, the collisional excitation rate is approximately balanced by the collisional de-excitation rate, such that the cooling rate scales approximately as $\Lambda_{\rm H_2}\propto n_{\rm H}$. The cooling time thus remains slightly larger than the free-fall time, and the collapse rate decreases.

The transition from non-LTE to LTE occurs approximately at a density of $n_{\rm H}\simeq n_{\rm H, crit}=10^4\,{\rm cm}^{-3}$, where the gas has cooled to $\simeq 200\,{\rm K}$. This phase in the collapse of the gas has been termed the `loitering phase', since the collapse momentarily slows while the central gas cloud continues to accrete mass \citep{bcl02}. The Jeans mass at this characteristic density and temperature is $\gtrsim 100\,{\rm M}_\odot$. Once this mass has been accreted, the cloud collapses under its own gravity. Since the H$_2$ cooling rate is a strong function of temperature, the cooling time $t_{\rm cool}$ in the LTE regime adjusts to the free-fall time. Equating these two time scales implies $\rho^{-1/2}\propto T^{1-\alpha}$, where $\alpha\simeq 4$ is the power-law exponent of the cooling function. Writing this in the form $T\propto\rho^{\gamma_{\rm eff}}$, where $\gamma_{\rm eff}$ is the effective equation of state, one finds $\gamma_{\rm eff}=7/6$. The temperature therefore increases very slowly with increasing density for $n_{\rm H}\gtrsim n_{\rm H, crit}$. In a near-isothermal, Jeans-unstable cloud the accretion rate is approximately given by ${\dot M}=M_{\rm J}/t_{\rm ff}\propto T^{3/2}$. Since the temperature of present-day star formation regions is about $10\,{\rm K}$, this simple physical argument implies that Population~III stars were typically $100$ times more massive than Population~I/II stars. 

\subsection{Three-body H$_2$ formation}

Following the nearly isothermal collapse of the cloud to densities $n_{\rm H}\simeq 10^8\,{\rm cm}^{-3}$, three-body reactions begin to convert the mostly atomic gas to molecular hydrogen \citep{pss83}. The most important formation reaction is
\begin{equation}
{\rm H}+{\rm H}+{\rm H}\rightarrow {\rm H}_2+{\rm H},
\end{equation}
with a smaller contribution from reactions involving H$_2$ molecules and helium atoms. The rapidly increasing H$_2$ fraction results in a similarly rapid increase in the H$_2$ line cooling rate. However, since each formation process is accompanied by the release of the binding energy of the H$_2$ molecule of $\simeq 4.48\,{\rm eV}$, the cooling is offset by chemical heating. The net effect is a mild drop in temperature as the H$_2$ fraction approaches unity. The resulting thermal instability of the gas caused by a chemical transition corresponds to the chemothermal instability envisaged by \citet{yoneyama73}. Previous studies have shown that it may trigger gravitational instability and fragmentation \citep{sy77, ra04, nu99, yoshida06b}. Indeed, in the simulations of \citet{tao09} and \citet{gsb13}, a subset of the clouds fragmented into two distinct clumps, although the strength of the instability in \citet{gsb13} was likely overestimated due to the approximate treatment of the optically thick H$_2$ line cooling rate (see Section~\ref{sec_line}). The further evolution of the clumps cannot be reliably extrapolated from the state of the cloud at the instant they fragment, but it appears that in some halos a wide binary system may form.

Until recently, one of the major caveats in the chemical evolution of primordial gas clouds was the large uncertainty in the three-body formation rate coefficient. At $1000\,{\rm K}$, the published rates differ by up to two orders of magnitude \citep{abn02, pss83, fh07, glover08}. The influence of the three-body formation rate on the properties of primordial gas clouds was investigated by \citet{turk11} and \citet{bsg14}. They found that radially averaged quantities as well as the morphologies of the clouds varied significantly between simulations with different rates, and are comparable to the differences for different initial conditions. This uncertainty may have recently been alleviated by the quantum-mechanical calculations of \citet{forrey13}. They found that the three-body formation rate coefficient at the relevant temperatures is approximately two times lower than that of the commonly employed \citet{glover08} rate.

\subsection{Optically thick H$_2$ line cooling} \label{sec_line}

Another important transition occurs at densities $n_{\rm H}\gtrsim 10^{10}\,{\rm cm}^{-3}$, where the gas becomes optically thick to H$_2$ line emission. For each of the $\simeq 200$ ro-vibrational transitions that are important at the relevant densities and temperatures, thermal Doppler broadening dominates the frequency dependence of the emitted radiation. The cross section for each line shows a similar frequency dependence, but in addition depends on the relative velocity along the line of sight. Until recently, this complicated radiative transfer problem has only been solved accurately in one-dimensional calculations \citep{omukai98, on98, ripamonti02}. These studies found that the wings of the lines allow the radiation to escape more efficiently than in the case of a grey opacity. In a spherically symmetric calculation, \citet{ripamonti02} and \citet{ra04} found that the escape fraction, which is defined as the probability of a photon to escape from the cloud, may be fit by $f_{\rm esc}={\rm min}\left[\left(n_{\rm H}/n_{\rm H, line}\right)^{-0.45}, 1\right]$, where $n_{\rm H,line}=6\times 10^9\,{\rm cm}^{-3}$.

This fit has also been used in three-dimensional simulations \citep{tao09, tna10, turk11, turk12}, due to the high computational cost of multi-frequency line transfer calculations in three dimensions. The accuracy of this treatment depends on how spherically symmetric the cloud is, and on how well the radially averaged profiles agree with those found in \citet{ripamonti02}. In particular, the functional form of the fit depends on the detailed chemical, thermal, and dynamical evolution of the cloud, which in turns depends on the various rate equations employed \citep{turk11}. According to the simulations of \citet{tao09}, the requirement of spherical symmetry is relatively well satisfied, while simulations that use the smoothed particle hydrodynamics method tend to find a more pronounced disk component \citep[e.g.][]{yoshida06b, clark11b}. The moving-mesh simulations of \citet{gsb13} found clouds with a morphology similar to those of \citet{tao09}. Only a few showed substantial deviations from spherical symmetry. In cases where these deviations exist, \citet{hy13} found that a simple density-dependent fitting function does not yield accurate results.

Another method to estimate the escape fraction is the Sobolev method \citep{sobolev60}. In a cloud with a uniform velocity gradient, the escape fraction is given by $f_{\rm esc}=\left[1-\exp{\left(-\tau\right)}\right]/\tau$, where $\tau=\alpha L_{\rm Sob}$, $L_{\rm Sob}$ is the Sobolev length, and $\alpha$ the absorption coefficient. The Sobolev length is given by $L_{\rm Sob}=v_{\rm therm}/\left|{\rm d}v_{\rm r}/{\rm d}r\right|$, where $v_{\rm therm}$ is the thermal velocity, and ${\rm d}v_{\rm r}/{\rm d}r$ the velocity gradient. The Sobolev length estimates the scale on which a line is Doppler-shifted out of resonance. This method was first used in the simulations of \citet{yoshida06b}, where the average escape fraction along the three principal axes of the computational domain was computed. In subsequent studies, this was replaced by the average velocity gradient, and the Sobolev length was limited to be smaller than the Jeans length for very small velocity gradients \citep{clark11a, clark11b, greif11a, greif12, gsb13}. In general, the Sobolev method is only valid if the Sobolev length is small compared to the scales on which the properties of the gas change significantly. This is not the case in primordial gas clouds, where transonic turbulence is present and the variation in the radial velocity is comparable to the sound speed of the gas. Nevertheless, \citet{turk11} found that the escape fractions obtained with the different methods disagree only by a factor of a few, while \citet{yoshida06b} and \citet{hy13} found agreement within a factor of two. However, the results of \citet{turk11} were obtained using different hydrodynamical schemes for the two methods, such that it is  unclear to what extent the discrepancy is related to the treatment of the optically thick cooling rate. 

\begin{figure}
\centering
\includegraphics[width=12cm]{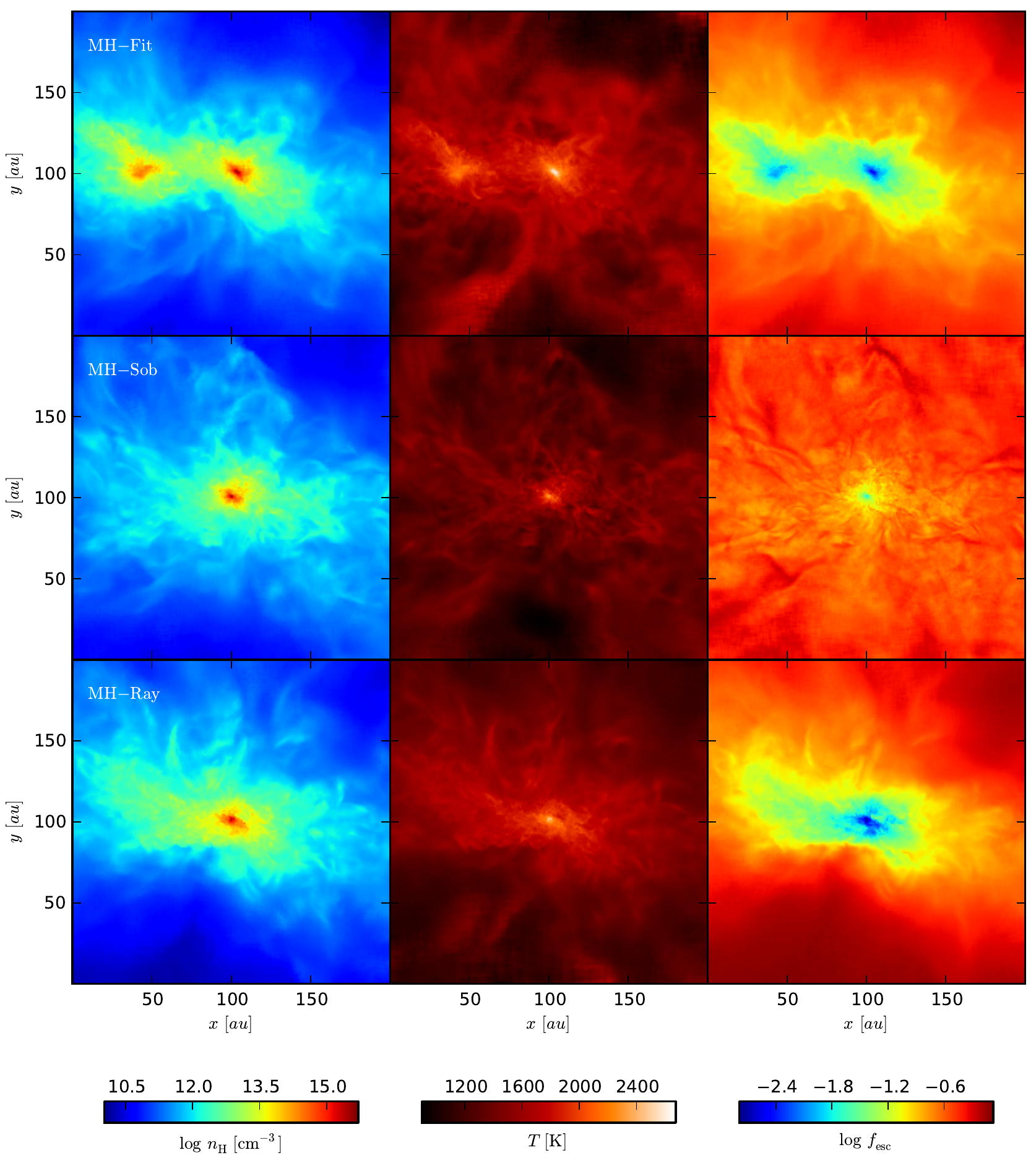}
\caption{Number density of hydrogen nuclei, temperature, and escape fraction of H$_2$ line photons in the central $200\,{\rm au}$ of a minihalo, shown for various treatments of the optically thick H$_2$ line cooling rate. The top row shows the outcome for the density-dependent fit of \citet{ra04}, the middle row for the Sobolov method, and the bottom row for the self-consistent radiative transfer calculation of \citet{greif14a}. The Sobolev method greatly overestimates the escape fraction at high densities, whereas the fitting function shows better agreement with the ray-tracing solution, even though the gas fragments in this particular simulation. Adapted from \citet{greif14a}.}
\label{fig_ray}
\end{figure}

The first self-consistent treatment of H$_2$ line transfer in a three dimensional simulation was presented in \citet{greif14a}. This study used a multi-line, multi-frequency ray-tracing scheme to follow the propagation of H$_2$ line radiation through the computational domain (see Figure~\ref{fig_ray}). The spherically averaged escape fraction agreed relatively well with the fit of \citet{ra04}, while the Sobolev method yielded escape fractions that were up to an order of magnitude higher. This discrepancy is due to the fact that the Sobolev approximation is not valid in primordial gas clouds. As a result, the gas temperature in \citet{greif14a} is systematically higher than in previous studies that used the Sobolev method, and the cooling instability found in \citet{gsb13} also disappears. The reduced temperature results in a somewhat reduced H$_2$ fraction at densities $n_{\rm H}\simeq 10^{15}\,{\rm cm}^{-3}$, which is in agreement with \citet{turk12}. An effect that is not captured in the previous approximate methods is the diffusion of radiation through the cloud, which tends to smooth out density perturbations. As a result, the central gas cloud becomes increasingly spherically symmetric as it collapses to higher densities. The fitting function also does not account for changes in the geometry of the cloud, which are expected once the collapse stalls and a disk forms. It neglects the complex dependence of the escape fraction on the properties of the gas, such as the density, velocity, and temperature profiles, and thus an implicit dependence on the various chemical and thermal rate equations employed.

Unfortunately, it is not yet computationally feasible to continue the radiation hydrodynamics simulations of \citet{greif14a} beyond the initial collapse of the gas \citep[e.g.][]{greif12}. This is in part due to the large number of opacity calculations necessary for an accurate integration, which is given by $N_\tau=N_{\rm cells}^{4/3}N_{\rm rays}N_{\rm lines}N_\nu$, where the individual factors denote the number of cells, the number of rays sent from each cell, the number of lines, and the number of frequency bins (the extra factor of $N_{\rm cells}^{1/3}$ accounts for the average number of cells an individual ray must traverse). In the simulation of \citet{greif14a}, these parameters were chosen to achieve an overall accuracy of $\simeq 5\%$. For $N_{\rm cells}\simeq 2\times 10^7$, $N_{\rm rays}=48$, $N_{\rm lines}=32$, and $N_\nu=8$, nearly $10^{14}$ opacity calculations per time step were carried out. A single ray-tracing step therefore took about $2000\,{\rm s}$ on $1024$ modern computing cores. To evolve the gas from $n_{\rm H}\simeq 10^{10}\,{\rm cm}^{-3}$ to $n_{\rm H}\simeq 10^{15}\,{\rm cm}^{-3}$, where the gas becomes optically thick to H$_2$ line emission, about $2000$ ray-tracing steps had to be performed. This corresponds to a total runtime of about one month and a computational cost of nearly one million CPU hours. The simulation was run using a hybrid MPI/Open-MP parallelization scheme on the Sandy-bridge cores of the supercomputer {\it Stampede} at the Texas Advanced Supercomputer (TACC). Among many other optimizations, the opacity calculation exploited the Intel AVX instruction set, which allows up to eight single-precision operations to be performed simultaneously.

Another challenge of the radiative transfer calculation is the amount of memory required to store the energy associated with the individual rays. In single precision, $4N_{\rm cell}N_{\rm rays}N_{\rm lines}N_\nu$ bytes must be reserved, which corresponds to $\simeq 1\,{\rm GB}$ per core for $1024$ cores. Each global communication step therefore requires nearly the entire memory to be exchanged between MPI tasks, which accounts for approximately $50\%$ of the total computational cost. Another problem is the significant imbalance in the number of rays stored on the tasks. Since the domain decomposition is optimized for hydrodynamical simulations instead of radiative transfer simulations, a significant imbalance accumulates as the rays are traversed. Typically, this amounts to another factor of two, and leads to a similar reduction in performance. It is therefore not yet beneficial to tailor the scheme for graphics processing units (GPU's) or co-processors.

An alternative method was recently introduced by \citet{hartwig14a}. This study used the {\sc treecol} algorithm of \citet{cgk12} to determine the total column density of H$_2$ molecules in various directions around each cell. The method accounts for relative velocities in a simplified manner, and computes the average photon escape fraction from the optical depth of the individual transitions. The computational cost is only about five times higher than in a simulation without radiative transfer, and allowed \citet{hartwig14a} to follow the build-up and fragmentation of the disk around the primary protostar. During the initial collapse, the escape fraction agrees relatively well with that of \citet{greif14a}, while in the later stages of the collapse the disk allows the radiation to escape more easily than previous approximate methods would imply. This reduces the thermal stability of the gas, and promotes fragmentation.

\subsection{Collision-induced emission and absorption}

\begin{figure}
\centering
\includegraphics[width=11cm]{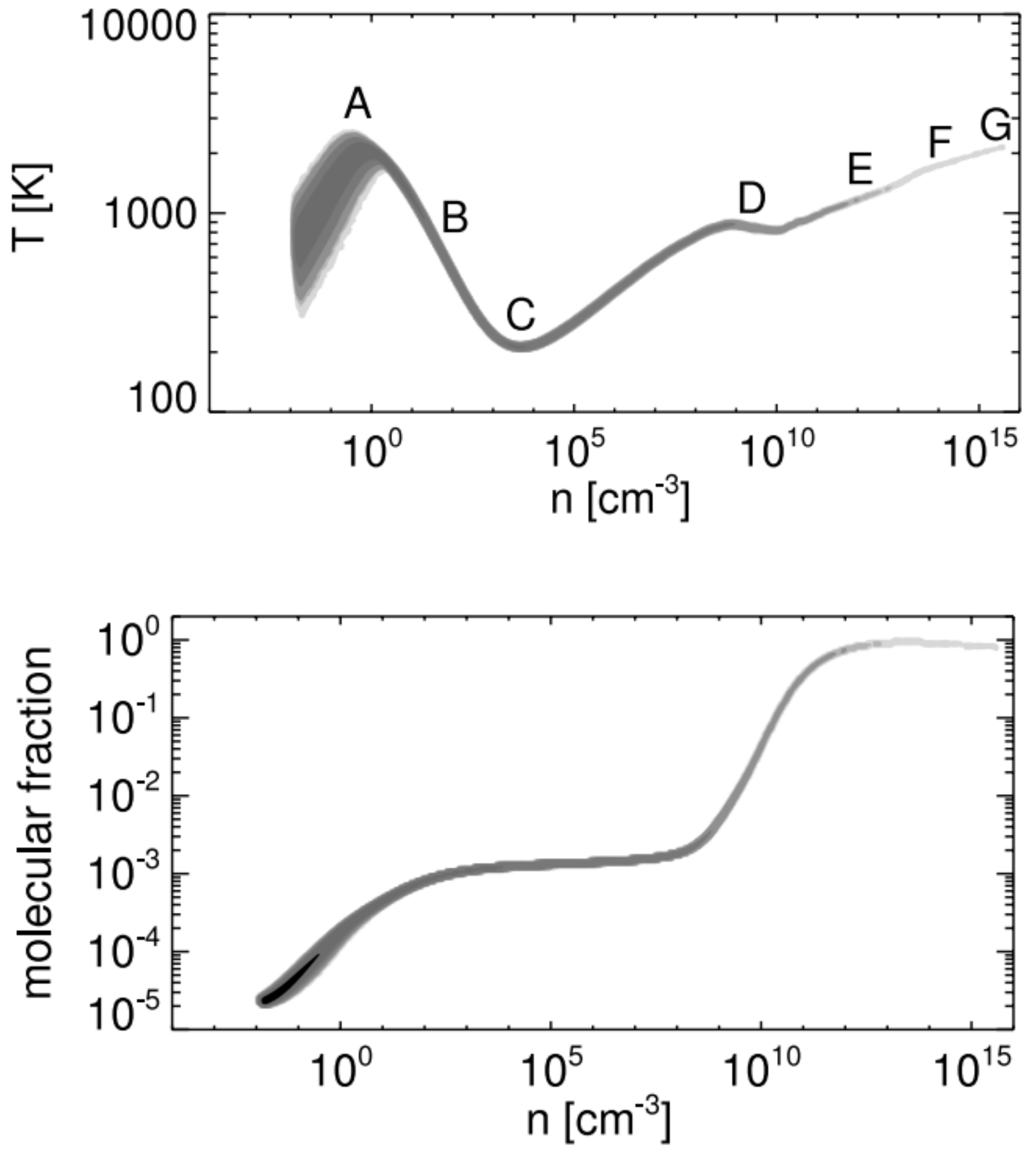}
\caption{{\it Top panel:} thermal evolution of primordial gas as it collapses over many orders of magnitude in density. The labels denotes various milestones in the collapse: heating to the virial temperature of the halo (A), runaway cooling via H$_2$ line emission (B), cooling to the minimum temperature of $\simeq 200\,{\rm K}$ (C), onset of three-body H$_2$ formation (D), gas becomes optically thick to H$_2$ line emission (E), onset of collision-induced emission (F), and collisional dissociation of H$_2$ (G). {\it Bottom panel:} H$_2$ fraction versus density. The H$_2$ fraction increases from its cosmological abundance of $\simeq 10^{-6}$ to $\simeq 10^{-3}$ via associative detachment of H and H$^-$. Following an extended plateau where the H$_2$ fraction remains nearly constant, the cloud becomes fully molecular once three-body reactions set in. Adapted from \citet{yoshida06b}.}
\label{fig_pspace}
\end{figure}

For densities $n_{\rm H}\gtrsim 10^{14}\,{\rm cm}^{-3}$, collision-induced emission and absorption become important \citep{on98, ripamonti02, ra04}. These processes are characterized by two hydrogen molecules approaching each other and forming a `super-molecule' with a dipole moment induced by van der Waals forces. The resulting super-molecule can emit or absorb radiation via dipole transitions, and has higher transition probabilities than the quadrupole transitions of isolated H$_2$ molecules. As opposed to H$_2$ line emission, where the line width is small compared to the separation of the individual lines, the short lifetime of the super-molecule results in line widths large enough that they merge into a continuum. \citet{yoshida06b} found that collision-induced emission continues to cool the gas even after it has become optically thick to H$_2$ line emission. The various stages in the collapse of the gas up to this point are shown in Figure~\ref{fig_pspace}.

In the one-dimensional dynamical model of \citet{ra04}, the gas becomes optically thick to collision-induced emission at densities $n_{\rm H}\simeq 10^{16}\,{\rm cm}^{-3}$. This rapid transition compared to H$_2$ line emission is due to the absence of absorption wings, which increases the optical depth of the gas. \citet{ra04} also found that a chemothermal instability similar in nature to the one induced by three-body H$_2$ formation may be triggered. However, the growth rate is longer than the free-fall time, such that it most likely does not result in sub-fragmentation of the cloud \citep[see also][]{yoshida06b}. In three-dimensional simulations, the continuum opacity of the gas has been taken into account by using the escape fraction method. The escape fraction is computed by evaluating the optical depth along the principal axes of the computational domain \citep{yoh08, hy13}, or by using a density-dependent fitting function \citep[e.g.][]{clark11a, greif11a}. The results achieved with these treatments can differ substantially for asymmetric cloud configurations, but agree relatively well with one-dimensional calculations for clouds that are approximately spherically symmetric \citep{hy13}.

\subsection{Collisional dissociation}

The last process that is able to cool the gas is collisional dissociation of H$_2$. It acts as a thermostat by converting the compressional energy of the gas into energy that is used for the dissociation of H$_2$, which has a binding energy of $4.48\,{\rm eV}$. The temperature and density at which H$_2$ begins to dissociate depends on the ratio of the three-body formation rate to the collisional dissociation rate. In \citet{tao09}, this occurs already at temperatures $\gtrsim 2000\,{\rm K}$ at a density of $n_{\rm H}\simeq 10^{15}$, while other studies find this to occur at significantly higher densities \citep{yoh08, clark11b, greif12}. This discrepancy is likely related to differences in the reaction rates and the treatment of the optically thick H$_2$ line cooling rate \citep{turk11, greif14a}.

From a computational standpoint, obtaining a self-consistent H$_2$ fraction at these densities is computationally expensive, since the three-body formation rate scales with the cube of the density. However, the H$_2$ abundance approaches chemical equilibrium at $n_{\rm H}\gtrsim 10^{15}\,{\rm cm}^{-3}$, and the electron abundance at $n_{\rm H}\gtrsim 10^{19}\,{\rm cm}^{-3}$. Once the density exceeds these critical values, an equilibrium solver instead of a non-equilibrium solver may be used, which greatly reduces the computational effort. Instead of evolving a system of stiff differential equations on a time scale that may be much smaller than the Courant time, the chemistry simplifies to a root-finding problem \citep{on98, ripamonti02, yoshida06b, yoh08, greif12}.

\section{First stars: the accretion phase} \label{sec_nsm}

\subsection{Protostar formation and evolution}

Following the dissociation of H$_2$, the temperature rises steeply with increasing density. At densities $n_{\rm H}\simeq 10^{20}\,{\rm cm}^{-3}$, the collapse stalls and an accretion shock forms that heats the gas to $\gtrsim 10^4\,{\rm K}$ \citep{on98, yoh08}. This marks the formation of a protostar at the center of the cloud with an initial mass of $\simeq 0.01\,{\rm M}_\odot$ and a size of $\simeq 0.1\,{\rm au}$.

The first calculations of the evolution of primordial protostars with continuous accretion were carried out by \citet{sps86b, sps86a} and \citet{op01, op03}. They solved the stellar structure equations for a hydrostatic core of radius $R_*$ bounded by an accretion shock with a constant accretion rate of about $5\times 10^{-3}\,{\rm M}_\odot\,{\rm yr}^{-1}$. A radiative precursor forms ahead of the accretion shock due to the high opacity of the gas, which may be many times larger than the protostar. Up to a mass of about $10\,{\rm M}_\odot$, the accretion time $t_{\rm acc}=M_*/\dot{M}_*$, where $M_*$ is the mass of the protostar, is smaller than the Kelvin-Helmholtz time $t_{\rm KH}=GM_*^2/\left(R_*L_*\right)$, where $L_*$ is the luminosity of the protostar. As a result, the protostar evolves nearly adiabatically and the radius increases to $\gtrsim 100\,{\rm R}_\odot$. Once $t_{\rm acc}>t_{\rm KH}$, this trend is reversed and the protostar undergoes Kelvin-Helmholtz contraction, which continues until $M_*\simeq 50\,{\rm M}_\odot$. Following a phase of deuterium burning, hydrogen burning begins after $\simeq 10^4\,{\rm yr}$, and the protostar settles onto the main sequence after $\simeq 10^5\,{\rm yr}$. Accretion is finally terminated when the luminosity becomes comparable to the Eddington luminosity, and the star has grown to a few hundred solar masses. These results agree with the more recent calculations of \citet{ho09} and \citet{ohkubo09}.

One of the major uncertainties of these models is the assumption of spherical symmetry. This was relaxed in \citet{tm04}, where a semi-analytic prescription for an accretion disk was employed. They found that the reduced density in the polar regions around the protostar results in a reduced optical depth, such that the photosphere of the spherically symmetric models disappears. This may have a substantial effect on the propagation of radiation from the protostar, and the maximum mass that can be reached.

\subsection{Accretion}

\citet{op03} showed that one of the main parameters that governs the evolution of protostars is the accretion rate. Neglecting radiative transfer effects and assuming that the gas does not fragment, the time-averaged accretion rate may be estimated as $\dot{M}\sim M_{\rm J}/t_{\rm ff}\sim c_{\rm s}^3/G\propto T^{3/2}$, where the Jeans mass is evaluated at the density and temperature when the gas first becomes gravitationally unstable, i.e. $n_{\rm H}\simeq 10^4\,{\rm cm}^{-3}$ and $T\simeq 200\,{\rm K}$. Since the temperature in present-day star-forming regions is $\simeq 10\,{\rm K}$, the accretion rate in primordial gas clouds is expected to be nearly $100$ times higher, with $\dot{M}\sim 10^{-3}\,{\rm M}_\odot\,{\rm yr}^{-1}$. Due to the Courant constraint, it is not possible to model the accretion phase accurately for more than a few free-fall times, or $\simeq 10\,{\rm yr}$ \citep{ripamonti02, greif12}. Since primordial stars are expected to accrete for more than $\sim 10^5\,{\rm yr}$ before radiation feedback terminates accretion \citep{tm04, mt08}, various methods have been used to estimate the final masses of Population~III stars. In \citet{on98}, the self-similar form of the spherically symmetric infall solution for $\gamma_{\rm eff}\simeq 1.1$ was extended to the accretion phase, resulting in $\dot{M}=8.3\times 10^{-2}\,{\rm M}_\odot\,{\rm yr}^{-1}\left(t/1\,{\rm yr}\right)^{-0.27}$, where $t$ denotes the time \citep{larson69, penston69, shu77, yahil83, ss83}. The mass of the star after $10^5\,{\rm yr}$ is thus $M_*\simeq 500\,{\rm M}_\odot$. A similar power-law for the accretion rate was found in the calculations of \citet{ripamonti02} and \citet{tm04}.

In three-dimensional simulations, the accretion rate may be obtained by measuring the spherically averaged density and velocity profiles \citep{abn02, yoshida06b, gao07, on07, turk11}. These instantaneous accretion rates generally agree with the power laws found in one-dimensional calculations. Studies that employed `sink particles' to represent growing protostars determined the accretion rate from the mass accreted by the sink particles \citep{bl04b, cgk08, sgb10, sbl11a, sgb12, sb13, clark11a, clark11b, greif11a, smith11, smith12a, sb14}. At sufficiently late times, they agree with those found in previous studies, but show substantial variability due to the instability and fragmentation of the accretion disk around the central protostar.

\subsection{Protostellar radiation}

\begin{figure}
\centering
\includegraphics[width=12cm]{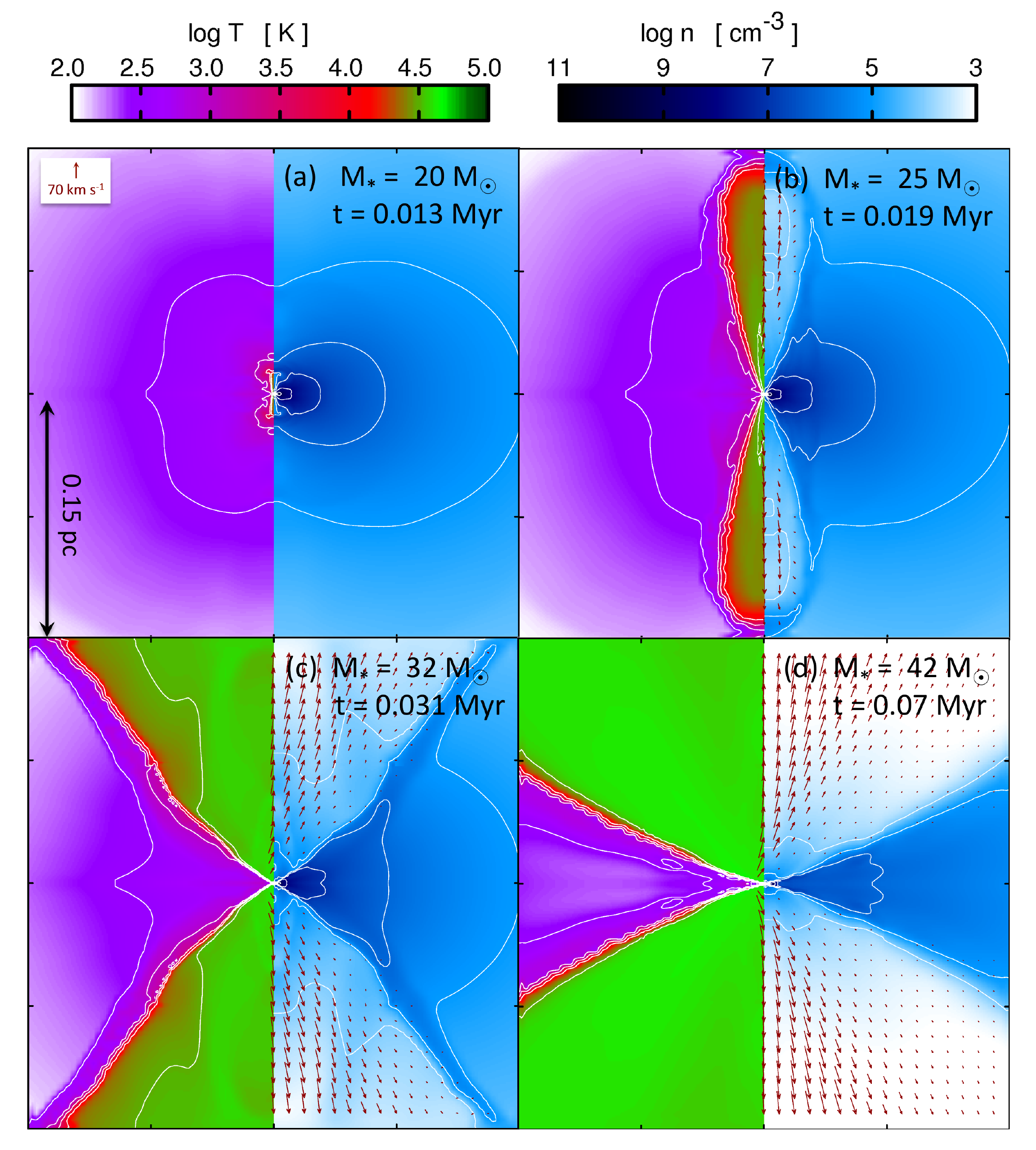}
\caption{Two-dimensional simulation of primordial star formation including radiation from the central protostar. The panels shows the temperature and number density of the cloud at various output times. The ionizing radiation clears out the gas along the poles, while the accretion continues nearly unabated in the plane of the disk. As the photo-heated region expands, it shuts down the mass flow from the envelope onto the disk, which eventually terminates accretion. Here, the star reaches a final mass of $\simeq 40\,{\rm M}_\odot$. Adapted from \citet{hosokawa11}.}
\label{fig_rad}
\end{figure}

Stellar radiation may shut down the accretion flow from the surrounding gas reservoir before the star reaches masses well in excess of $100\,{\rm M}_\odot$. Since the influence of both stellar winds and radiation pressure are expected to be small in the absence of metals and dust grains \citep{bhw01, bkl01, marigo01, kudritzki02, oi02, krticka03, mck03, kk06, mt08, kk09, kvk10, ss11, muijres12, ss12, su12}, the most important feedback mechanisms are the dissociation of H$_2$ and the heating that accompanies the ionization of neutral hydrogen atoms. The former occurs due to radiation in the so-called Lyman-Werner (LW) bands, which indirectly dissociates H$_2$ and removes the most important coolant of the gas \citep{on99, gb01}. \citet{hosokawa11} argued that this process is not particularly important, since the disk remains optically thick to LW radiation, while the molecules in the bipolar regions perpendicular to the disk are collisionally dissociated by photoheated gas (see Figure~\ref{fig_rad}).

Once the star undergoes Kelvin-Helmholtz contraction, the effective temperature becomes high enough for significant amounts of ionizing radiation to be produced. The energy above $\simeq 13.6\,{\rm eV}$ thermalizes and heats the gas to $\gtrsim 10^4\,{\rm K}$, which is significantly higher than the virial temperature of a minihalo. The high pressure therefore begins to drive gas from the halo. However, in the one-dimensional calculations of \citet{oi02}, the nascent H\,{\sc ii} region remains compact due to the high inflow velocity of the gas, and does not impede the accretion flow. The accretion disk model of \citet{mt08}, on the other hand, predicts a reduced density and inflow velocity along the poles, which allows the ionizing radiation to escape and begin to photo-evaporate the disk. This is confirmed by the two-dimensional simulations of \citet{hosokawa11} and the three-dimensional simulations of \citet{sgb12}, which used ray-tracing methods to compute the propagation of the radiation from the protostars. In the case of \citet{hosokawa11}, flux-limited diffusion was used to model the diffuse component of the radiation, and the effects of radiation pressure were included. These studies found an upper mass limit of a few tens of solar masses.

\begin{figure}
\centering
\includegraphics[width=12cm]{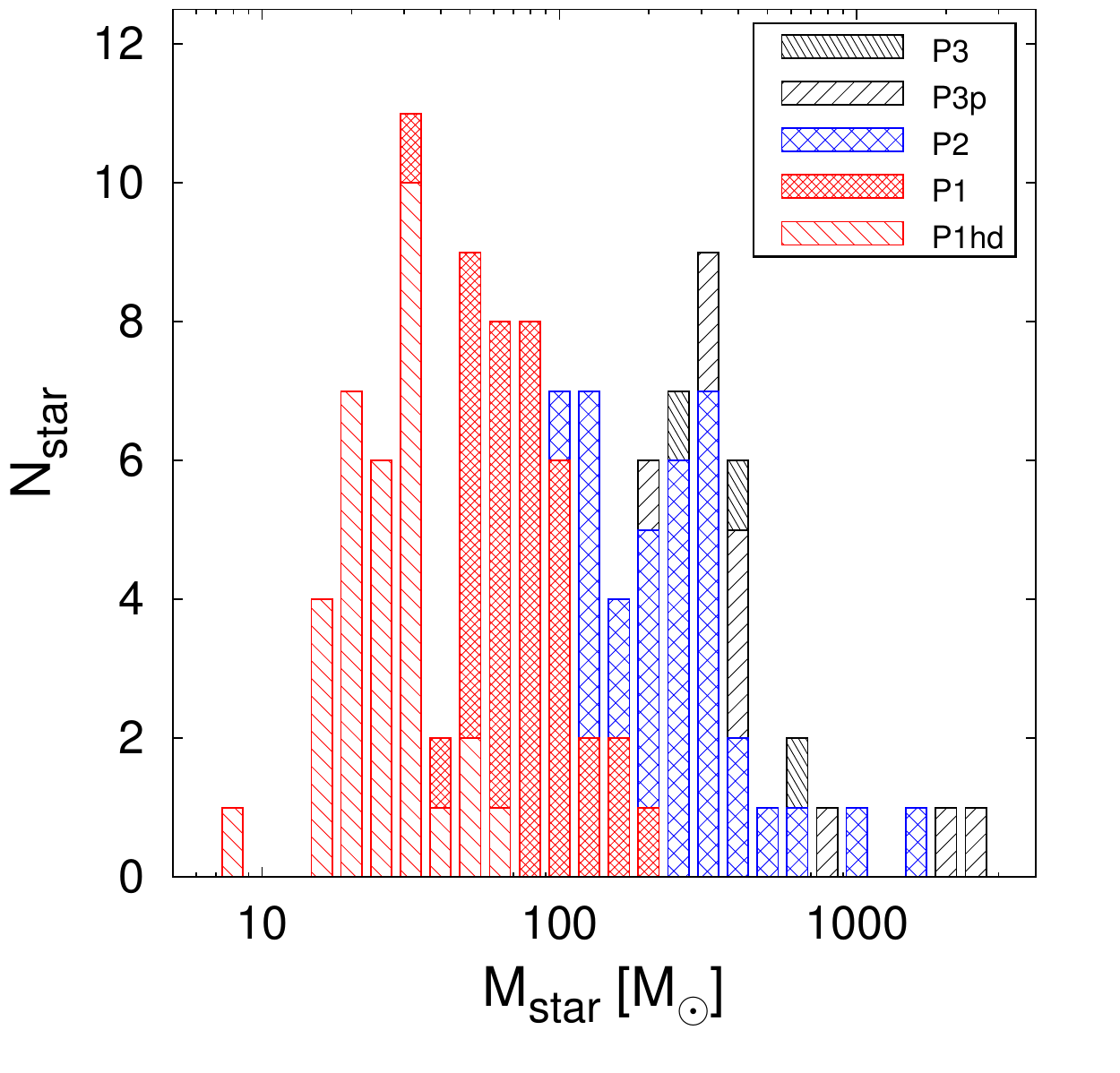}
\caption{IMF of Population~III according to the two-dimensional radiation hydrodynamics simulations of \citet{hosokawa11}, but applied to $100$ different minihalos. The various colors and hatchings represent different paths of protostellar evolution: P1 denotes a Kelvin-Helmholtz contracting protostar, P1hd a halo in which HD cooling was activated, P2 an oscillating protostar, and P3 and P3p super-giant protostars. Adapted from \citet{hirano14}.}
\label{fig_imf}
\end{figure}

Recently, \citet{hirano14} used the methodology of \citet{hosokawa11} to investigate the influence of radiation in a sample of $100$ different minihalos that were extracted from three-dimensional cosmological simulations. They found final stellar masses in the range $\simeq 10$--$1000\,{\rm M}_\odot$, which are correlated with the thermal evolution of the gas during the initial collapse phase (see Figure~\ref{fig_imf}). In contrast to these studies, \citet{susa13} and \citet{sht14} found that the molecule-dissociating radiation from the central protostar is more important and indirectly shuts down accretion. However, their resolution was not high enough to resolve the ionization front along the poles. They found that most Population~III stars had final masses in the range $10\lesssim M_*\lesssim 100\,{\rm M}_\odot$.

Despite the significant progress made by these studies, some caveats remain. The simulations of \citet{hosokawa11} and \citet{hirano14} included the most detailed treatment of radiative transfer, but were limited to two dimensions. The three-dimensional simulations of \citet{sgb12}, \citet{susa13} and \citet{sht14}, on the other hand, suffered from limited resolution and a simplified treatment of the radiative transfer. Significant progress could be made if the physical detail of the simulations of \citet{hosokawa11} were applied to three dimensions.

\subsection{Fragmentation}

Fragmentation during the accretion phase may further reduce the mass of the central protostar. If the cloud fragments shortly after the formation of the first protostar, much of the material in the surrounding envelope may be accreted by secondary protostars before they reach the center of the cloud. This process has been termed fragmentation-induced starvation \citep{peters10b}. Although fragmentation may occur already during the initial collapse phase \citep{tao09, gsb13}, most of the fragmentation is expected to occur after the first protostar has formed. Due to the Courant constraint, numerical simulations are not able to self-consistently evolve the collapse of the gas for a significant period of time beyond the initial collapse \citep{on98, ripamonti02, greif12}. For this reason, sink particles have been used to represent growing protostars and circumvent the need to model their interior structure and evolution \citep{bbp95,kmk04,jappsen05,federrath10}.

Next to criteria involving the thermal, kinetic, and gravitational binding energy of the gas, as well as the properties of the local velocity field, sink particles are typically inserted at a threshold density and accrete gas within a predefined accretion radius. This prevents the gas from collapsing to increasingly high densities, and allows the the simulations to be continued for a much longer period of time than would otherwise be possible. By nature, the sink particle method has its own limitations. The boundary conditions imposed on the gas near the accretion radius are necessarily artificial and may lead to unphysical results. The complicated torques on the scale of the accretion radius are not captured accurately, which might artificially enhance or prevent fragmentation. The loss of resolution on the scale of the accretion radius also sets a minimum scale on which fragmentation can occur. Close encounters between sink particles may result in dynamical ejections, although the orbital energy may in fact be dissipated through unresolved torques. Nevertheless, the sink-particle technique is currently the only method to investigate the evolution of primordial gas clouds over significant periods of time, and efforts are underway to increase their physical realism \citep{hww13}.

Some of the first studies that employed sink particles inserted them on scales significantly larger than the expected sizes of the protostars \citep{bcl02, bl04b, sgb10, sb13}. It was therefore not clear whether the gas represented by the sink particles would further sub-fragment. \citet{cgk08} addressed this problem by using a threshold density of $n_{\rm H}=10^{16}\,{\rm cm}^{-3}$, which corresponds to a radius of $\simeq 1\,{\rm au}$ and is the approximate size of a metal-free protostar that has a realistic accretion rate \citep{sps86b, sps86a}. The initial conditions were designed to reproduce the end state of the primordial gas clouds found in \citet{bcl02}, and they used the tabulated equations-of-state of \citet{omukai05} for various metallicities instead of a chemical network. In the primordial case, the central gas cloud fragmented into $25$ protostars after only a few hundred years, with masses ranging from $\simeq 0.02$ to $\simeq 10\,{\rm M}_\odot$. As opposed to the simulations with a non-negligible metallicity, the mass function in the primordial case was relatively flat, implying that most of the mass is locked up in high-mass protostars. In a subsequent study, \citet{clark11a} improved upon their previous work by using a detailed chemistry and cooling network. They investigated how the fragmentation depends on the mean Mach number of the turbulence, and found that a higher degree of turbulence generally leads to more fragmentation.

\begin{figure}
\centering
\includegraphics[width=13cm]{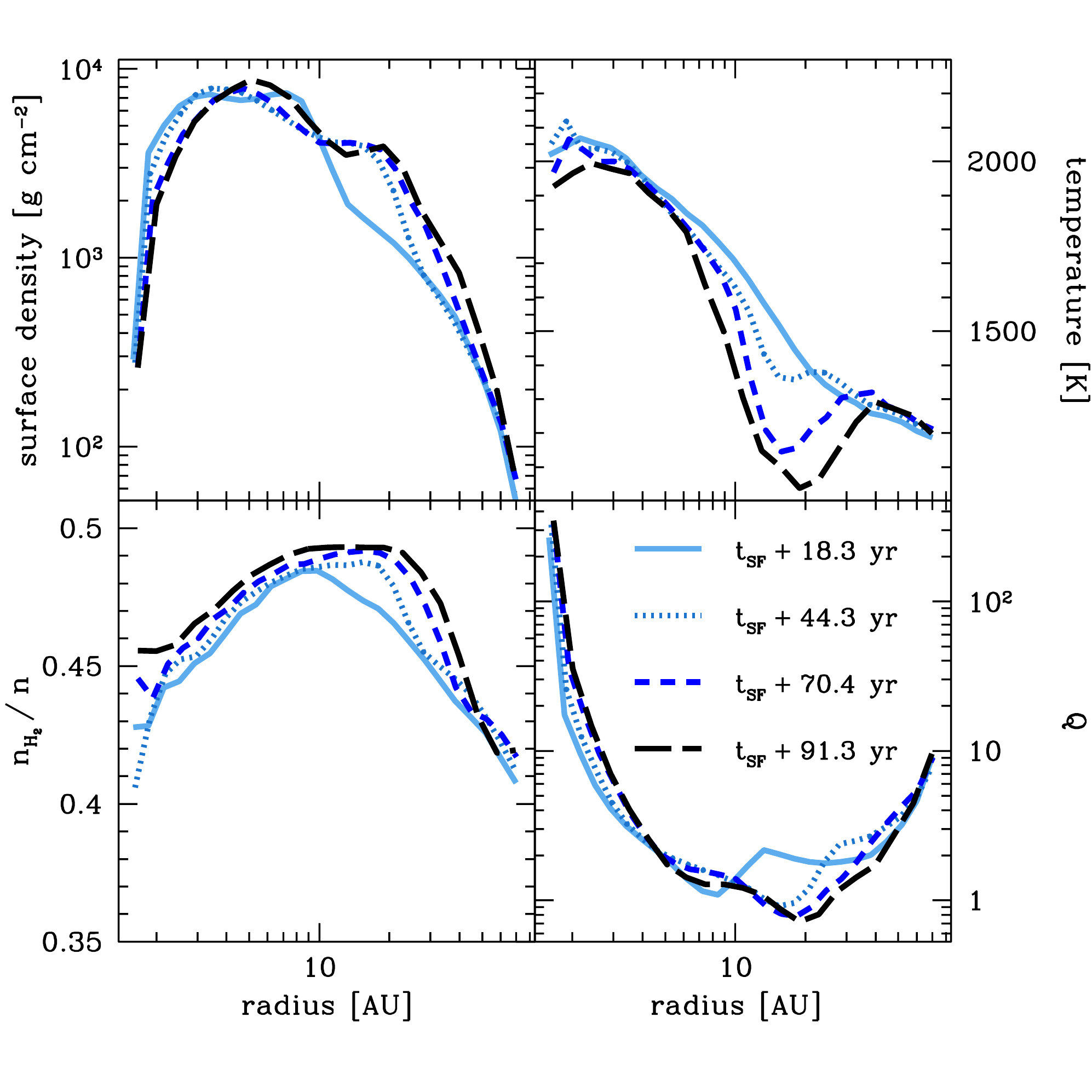}
\caption{{\it From top left to bottom right:} Surface density, temperature, H$_2$ fraction, and Toomre $Q$ parameter in the Keplerian disk that forms around the central protostar. For $Q\lesssim 1$, the disk becomes gravitationally unstable and fragments. In this simulation, the first fragment forms after $\simeq 90\,{\rm yr}$ at a distance of $\simeq 20\,{\rm au}$ from the central protostar. Adapted from \citet{clark11b}.}
\label{fig_toomre}
\end{figure}

The remaining caveat of idealized initial conditions was addressed in \citet{clark11b}. They employed cosmological initial conditions and a hierarchical zoom-in procedure to extract the central, Jeans-unstable cloud \citep{nw94, tbw97, gao05}. The resolution was increased several times with a particle-splitting technique \citep{kw02, bl03b}. This allowed the density at which sink particles were created to be increased to $n_{\rm H}=10^{17}\,{\rm cm}^{-3}$. \citet{clark11b} also included the effects of radiative heating due to accretion using a Planck mean opacity, based on the assumption that the gas was optically thin. They found that an accretion disk forms around the central protostar, which develops pronounced spiral arms. The first fragments appear after about $100\,{\rm yr}$, and by the end of the simulation three additional fragments have formed. The fragmentation is attributed to the limited rate at which mass can be processed through the disk, which is significantly smaller than the accretion rate from the surrounding envelope onto the disk. The surface density of the disk increases, while the compressional energy is radiated away by H$_2$ line emission. This quickly drives the disk towards gravitational instability (see Figure~\ref{fig_toomre}). Previous semi-analytic studies found that the disk remains stable, since they assumed that H$_2$ cooling is inefficient at the densities and temperature at which the disk forms \citep{tm04, tb04, md05a}. \citet{to14} found that primordial disks remain marginally stable despite the H$_2$ cooling \citep[but see][]{ls14a, ls14b}. It is possible that the ability of the gas to cool may have been overestimated in previous studies that employed the Sobolev method to compute the optically thick H$_2$ line cooling rate \citep[see][]{greif14a}. However, the more accurate method used in \citet{hartwig14a} showed that this does not substantially affect the fragmentation of the gas.

The simulations of \citet{greif11a} were similar in nature to those of \citet{clark11a}, but employed a moving-mesh \citep{springel10a} instead of an SPH approach \citep{syw01, springel05a}. Using the same chemical model and a similar sink-particle scheme, \citet{greif11a} investigated fragmentation in five different minihalos over a ten times longer period of time than in \citet{clark11a}. They found that on average ten protostars per halo had formed after $\simeq 10^3\,{\rm yr}$, with masses ranging from $\simeq 0.1$ to $\simeq 10\,{\rm M}_\odot$. The mass function was similarly flat as in \citet{cgk08}. \citet{greif11a} also found that a number of protostars with masses below $0.8\,{\rm M}_\odot$ formed, which would allow them to survive to the present day. These protostars were ejected from the central gas cloud by N-body interactions with more massive protostars and stopped accreting. In some cases, their velocities exceeded the escape velocity of the halo. However, the fraction of ejected protostars also depended strongly on the implementation of the sink-particle method, since the gas surrounding the protostars may absorb a significant fraction of their orbital energy.

The effects of radiative heating by the protostars was investigated by \citet{smith11} and \citet{smith12a}, using the halos of \citet{greif11a} and the radiative transfer method employed in \citet{clark11b}. Due to the low effective temperature of the protostars during the adiabatic expansion phase, most of the radiation is in the infrared and therefore serves to only slightly heat the gas. Fragmentation occurs somewhat later than in \citet{greif11a}, and the number of fragments that form is somewhat reduced.

\subsection{Protostellar system}

In an attempt to avoid the uncertainties introduced by sink particles, \citet{greif12} self-consistently evolved the collapse of the gas beyond the formation of the first protostar. Due to the substantial computational effort involved in resolving the Jeans length with $32$ cells at all times, only the first $\simeq 10\,{\rm yr}$ in the evolution of the protostellar disk could be modeled. Five different halos with the same initial conditions as in \citet{greif11a} were investigated. In analogy to previous studies, a Keplerian disk formed that developed spiral arms and fragmented. The gravitational stability of the disk was evaluated using the so-called \citet{toomre64} and \citet{gammie01} criteria. The Toomre $Q$ parameter, which quantifies the stability of the disk to perturbations, is given by $Q=c_{\rm s}\kappa/\left(\pi G\Sigma\right)$. Here, $\kappa$ is the epicyclic frequency of the perturbation, which is equal to the angular velocity $\Omega$ in a Keplerian disk, and $\Sigma$ is the surface density. For $Q<1$, perturbations in the disk can grow, but these perturbations do not necessarily result in fragmentation. The latter requires the Gammie criterion, $t_{\rm cool}<3/\Omega$, where $t_{\rm cool}$ is the cooling time, to be satisfied. \citet{greif12} showed that both of these conditions are fulfilled on a scale of $\simeq 1\,{\rm au}$, which agrees with the location of the fragmentation in the simulations. The efficient cooling of the gas necessary to fulfill Gammie's criterion stems mainly from the dissociation of H$_2$, which acts as a thermostat by extracting compressional energy from the gas and using it to unbind H$_2$. In this respect the results of \citet{greif12} differ somewhat from \citet{clark11a}, where H$_2$ line emission was found to be the most important coolant. This is most likely due to the higher resolution employed in \citet{greif12}, which allowed the collapse of the gas to be followed to significantly higher densities, where collisional dissociation cooling begins to dominate over H$_2$ line emission.

The dependence of the stability of the disk on the abundance of H$_2$ implies that differences in the three-body formation rates and the treatment of the optically thick H$_2$ line cooling rate may significantly affect the fragmentation of the gas \citep[e.g.][]{turk12}. The recent simulations of \citet{greif14a} show that the ability of the gas to cool may indeed have been overestimated. However, the strong asymmetry of the cloud that develops over time should allow the radiation to escape more easily than in the spherically symmetric model for the optically thick H$_2$ line cooling rate used in \citet{turk12}. Indeed, this trend has been confirmed by the simulations of \citet{hartwig14a}. In the future, it will be useful to evolve the self-consistent simulations of \citet{greif14a} to higher densities, but with modifications to the radiative transfer scheme that make them computationally feasible.

\begin{figure}
\centering
\includegraphics[width=13cm]{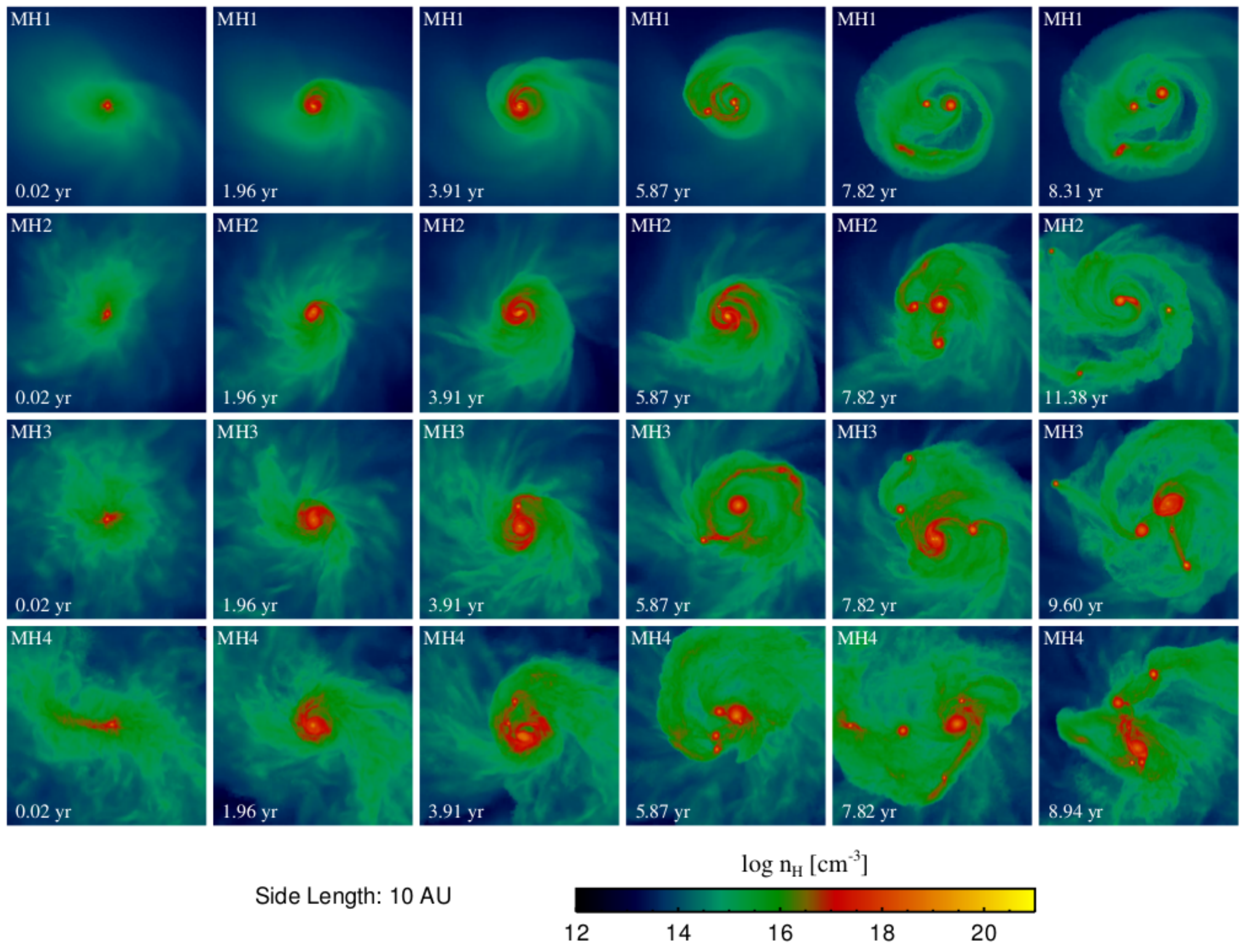}
\caption{{\it From left to right:} the first $\simeq 10\,{\rm yr}$ of the evolution of the protostellar system that forms at the center of four different minihalos. In all cases, a disk forms that becomes gravitationally unstable and fragments into a small system of protostars. Complicated gravitational torques result in a fraction of the protostars migrating to the center of the cloud, where they merge with the primary protostar, while others migrate to higher orbits. Adapted from \citet{greif12}.}
\label{fig_proto}
\end{figure}

The explicit resolution of the interface between the protostars and the disk in \citet{greif12} allowed the complex interactions of the protostars with other protostars and the surrounding gas to be modeled self-consistently (see Figure~\ref{fig_proto}). This study found that the protostars are subject to strong gravitational torques that lead to the migration of about half of the secondary protostars formed in the disk to the center of the cloud, where they merge with the primary protostar. The aggressive migration and merging of the protostars occurs on a free-fall time scale. In analogy to \citet{greif11a}, some low-mass protostars migrate to higher orbits due to N-body interactions, but do not become nearly as unbound as in \citet{greif11a}. This is mainly because the protostars have extremely large radii of the order of $100\,{\rm R}_\odot$, such that protostellar interactions do not resemble those of the point masses in \citet{greif11a}. Much of the orbital energy is transferred to the surrounding gas instead of the protostars. At the end of the simulation, about five protostars per halo are present, with a tendency to further increase. The mass budget is dominated by the primary protostar, which grows to about five times the mass of all other protostars.

In a qualitatively similar study, \citet{latif13c} artificially truncated the collapse of the gas at a density of $n_{\rm H}\simeq 10^{13}\,{\rm cm}^{-3}$. Even though the size of the disk was much larger than in \citet{greif12}, they found a similar susceptibility to fragmentation. \citet{vdb13} employed two-dimensional simulations with a predefined effective equation-of-state to investigate the evolution of a metal-free protostellar system for $\simeq 5\times 10^4\,{\rm yr}$. In this study, the secondary protostars quickly merged with the primary protostar, but the disk continued producing new fragments, with of order $10$ protostars present at any given time.

Despite the limitations of these simulations, they demonstrate the advantages and disadvantages of the sink-particle technique. Sink particles allow the simulations to be continued for much longer than would otherwise be possible, and probe the chemical and thermal evolution of the gas on scales larger than the accretion radius. On the other hand, they are not well suited to predict the properties of the protostars themselves, which depend sensitively on the properties of the gas on scales comparable to or smaller than the accretion radius. Since self-consistent simulations are extremely expensive, sink particles remain the only method capable of probing sufficiently late times in the Population~III star formation process.

\subsection{Stellar rotation}

Early simulations showed that the cloud out of which the first protostellar seed forms tends to also have a strong rotational component \citep[e.g.][]{abn02, bcl02}. Due to turbulence and other instabilities, the angular momentum is constantly redistributed as the gas continues to collapse, maintaining rotational velocities that are a significant fraction of the Keplerian velocity \citep[e.g.][]{yoshida06a, greif12}. The protostar that forms at the center of the cloud is therefore endowed with significant rotation. Since it accretes material from a Keplerian disk, the angular momentum of the protostar may remain high as it evolves towards the main sequence. In simulations of present-day star formation, magnetic braking may reduce the rotation rate of the protostar, but their influence in the primordial case is not yet clear. Rapidly rotating Population~III stars can have very different properties than their non-rotating counterparts, since rotation affects their evolution on the main sequence and beyond, as well as the degree to which the nucleosynthetic products mix with each other. The amount of rotation may also affect the types and properties of their SN explosions (see Section~\ref{sec_sn}).

\citet{sbl11} employed sink particles that traced the angular momentum of the accreted gas to investigate the rotation rate of primordial protostars. They found that the protostar represented by the sink particle rotated at near break-up speeds throughout the $\simeq 5000\,{\rm yr}$ that they simulated. In a follow-up study, \citet{stacy13} analyzed the rotation of the protostars formed in the simulations of \citet{greif12}, which self-consistently modeled the interface between the protostars and the surrounding gas. Similar to \citet{sbl11}, they found that the central protostar rotated at a significant fraction of the Keplerian velocity, despite strong gravitational torques due to frequent mergers with secondary protostars. Stellar evolution models should therefore account for the rotation of the star.

\section{First stars: additional physics} \label{sec_ap}

\subsection{HD cooling} \label{sec_hd}

The second low-temperature coolant in primordial gas that may become important is hydrogen deuteride \citep[HD;][]{ls84, puy93, sld98, gp98, gp02, fr99, flower00a, flower00b, fp01, lna05, jb06, ga08}. As opposed to H$_2$, HD possesses a permanent dipole moment, with correspondingly higher radiative transition probabilities. This shifts the critical density at which the level populations transition to LTE to $n_{\rm H, crit}\sim 10^6\,{\rm cm}^{-3}$. Furthermore, since ortho and para states do not exist, the transition $J=1\rightarrow 0$ is accessible, corresponding to an excitation temperature of $\simeq 130\,{\rm K}$. The primary formation reaction is:
\begin{equation} \label{eq_hd}
{\rm H}_2 +{\rm D}^+\rightarrow {\rm HD}+{\rm H}^+,
\end{equation}
showing that the HD formation rate depends on the ionization state of the gas and the abundance of H$_2$. Since HD is primarily destroyed by the inverse reaction of equation~\ref{eq_hd}, the difference in the zero-point energies of H$_2$ and HD can lead to a significant degree of chemical fractionation at low temperatures, which boosts the HD/H$_2$ ratio well above the cosmological D/H ratio of about $10^{-5}$.

In minihalos, the temperature usually does not become low enough for HD cooling to become important \citep{bcl02}. However, more recent studies have found that HD cooling may in fact dominate over H$_2$ cooling in certain halos. \citet{ripamonti07} and \citet{mb08} showed that HD cooling is activated in high-redshift minihalos with masses below $\simeq 3\times 10^5\,{\rm M}_\odot$. This allows the gas to cool down to the temperature of the CMB, which lies between $\simeq 50$ and $\simeq 100\,{\rm K}$ at $z\gtrsim 20$. The corresponding Jeans mass is up to an order of magnitude lower than in the pure H$_2$ cooling case, and may lead to the formation of a distinct population of metal-free stars with a characteristic mass of $10\,{\rm M}_\odot$ \citep[see also][]{nu02}.

A similar decrease of the initial Jeans mass was found in some of the halos investigated in \citet{greif11a} and \citet{gsb13}. In the radiation hydrodynamics simulations of \citet{hosokawa12} and \citet{hirano14}, the lower accretion rates in halos in which HD cooling was activated led to final stellar masses $\gtrsim 10\,{\rm M}_\odot$. \citet{clark11a} came to a somewhat different conclusion. They found that the gas in these clouds heated more rapidly at densities $n_{\rm H}\gtrsim n_{\rm H, crit}$, which suppresses turbulence and results in less fragmentation. The accreted mass is therefore distributed to fewer protostars, which allows them to become more massive. The influence of fragmentation on the growth of the protostars could not be addressed in the two-dimensional simulations of \citet{hosokawa12} and \citet{hirano14}, while \citet{clark11a} evolved the central cloud for only a few thousand years, and did not model the radiation from the protostars. It is therefore not yet clear whether HD cooling acts to increase or decrease the typical mass of Population~III stars. However, the total mass in stars is likely lower in the HD cooling case. The number of minihalos in which HD cooling is activated may be significantly reduced by a global LW background, which dissociates H$_2$ and thus reduces the rate at which HD forms \citep{wh11}.

If the electron abundance is significantly enhanced with respect to the post-recombination value, HD cooling may become important in other circumstances as well. An elevated electron abundance is produced by the virialization shocks of atomic cooling halos or during mergers \citep{gb06, greif08, sv06, vs08, pjm12, bovino14a, pjv14}, as well as in SN remnants \citep{ms86, ui00, mbh03, machida05, vs05b, vvs08}. The H$_2$ abundance in the post-shock gas increases to values well above those found in minihalos, which allows the gas to cool to temperatures low enough that chemical fractionation occurs and HD cooling takes over. A similar process occurs in relic H\,{\sc ii} regions, where the recombination time is long enough to facilitate the formation of HD \citep{no05, jb07, yoshida07, yoh07, mom09}. A distinct population of metal-free stars may thus arise under circumstances where the gas has been affected by radiation, but is not yet enriched with metals.

\subsection{Magnetic Fields}

Magnetic fields have been neglected in most studies of primordial star formation, even though it has been shown that they are important in protogalactic halos \citep{ps89, beck94, lc95, kulsrud97}. The magnetic field strength at the time of the formation of the first stars is constrained to $B\lesssim 1\,{\rm nG}$ \citep{sbk08}, although they are likely much weaker. Potential seed fields are generated during inflation, the electroweak phase transition, and the quark-hadron phase transition \cite[for a recent review, see][]{widrow12}. A more robust formation mechanism is the Biermann battery \citep{biermann50}, which requires the density gradient in an ionized gas to be misaligned with the pressure gradient, and is thus coupled to the vorticity of the gas. \cite{xu08} investigated the growth of the magnetic field in a three-dimensional simulation of a minihalo via the Biermann battery, and found that it generates a seed field of the order of $10^{-18}\,{\rm G}$ \citep[see also][]{ds11}. Another possible formation mechanism at a later stage in the collapse of the gas is via radiative forces \citep[e.g.][]{ssh14}.

\begin{figure}
\centering
\includegraphics[width=13cm]{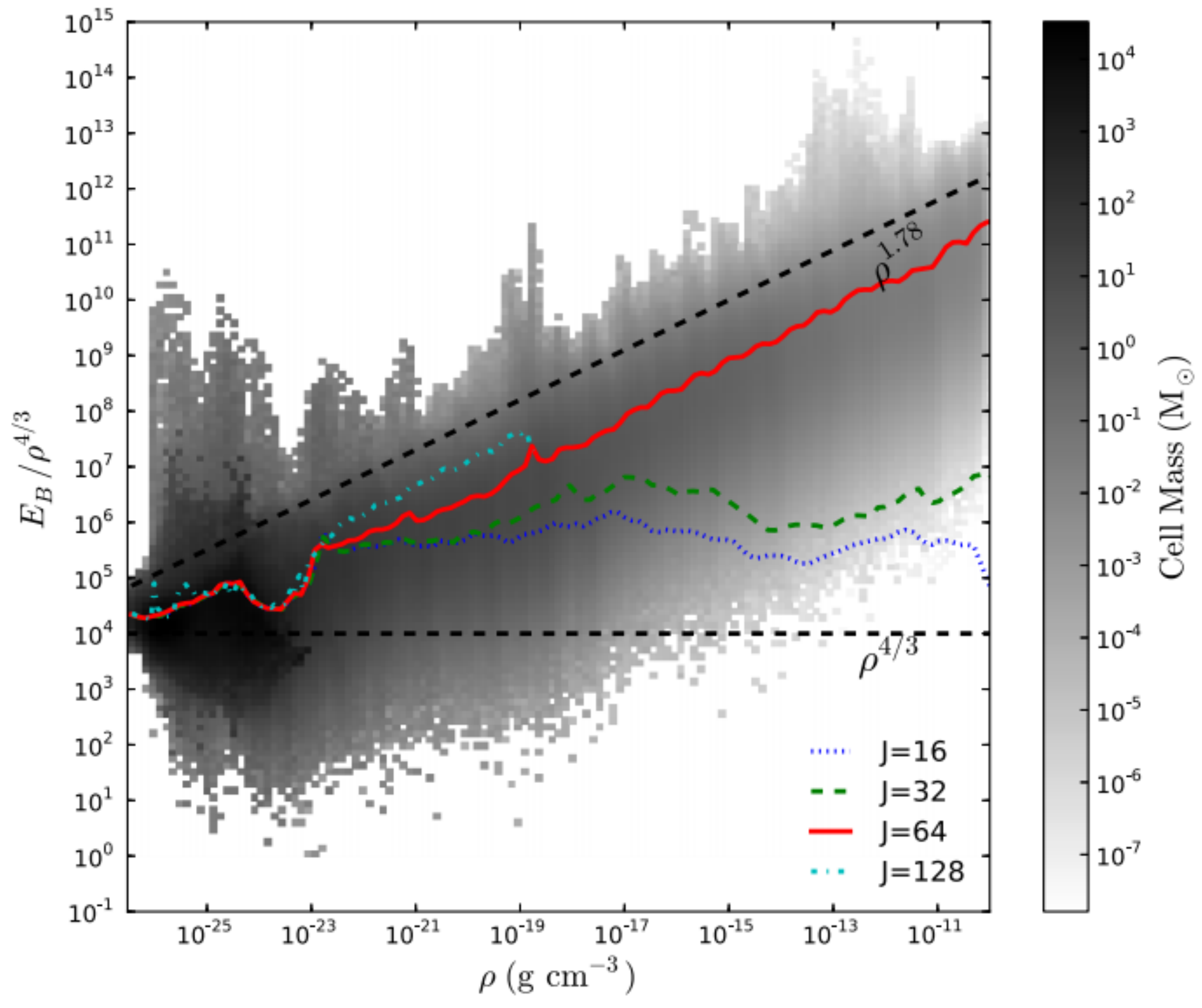}
\caption{Volume-averaged magnetic energy scaled by $\rho^{4/3}$ as a function of density in a simulation of primordial star formation that includes magnetic fields. The various line styles denote the resolution of the Jeans length. If the magnetic flux is frozen into the flow, the magnetic energy evolves along the lower black dashed line. Depending on resolution, the turbulent dynamo amplifies the magnetic field strength well above this value. Here, at least $64$ cells per Jeans length are required for significant amplification to occur. The shaded regions show the distribution of the gas in the $J=64$ case. There is no indication that the magnetic energy converges with increasing resolution. Magnetic fields will likely become dynamically important during the initial collapse, and affect the susceptibility of the gas to fragmentation. Adapted from \citet{turk12}.}
\label{fig_mag}
\end{figure}

If the time scale for ambipolar or Ohmic diffusion is large compared to the evolutionary time of a system, the magnetic field is `frozen' into the gas and moves with it \citep[but see][]{ms04, ms07}. In a spherically symmetric, contracting gas cloud the magnetic field grows $\propto\rho^{2/3}$, while more asymmetric configurations lead to a shallower power-law. For a gas in which the temperature does not evolve substantially, the ratio of the thermal pressure to the magnetic pressure thus decreases $\propto \rho^{-1/3}$, such that a seed field generated by the Biermann battery and amplified by flux-frozen collapse does not become dynamically important. A much more potent amplification mechanism is the turbulent dynamo \citep{parker63, kn67}. Small-scale turbulent motions in the gas repeatedly fold the magnetic field, which can grow by many e-foldings in a free-fall time. Simple calculations have shown that this mechanism can amplify the magnetic field in primordial gas clouds to appreciable levels \citep{schleicher10, schober12}. This was confirmed by three-dimensional simulations that included the effects of magnetic fields \citep{sur10, peters12, latif13b, lss14}. These studies also found that the Jeans length must be resolved by at least $32$ cells for significant amplification to occur \citep[see also][]{federrath11}, and that the amplification rate does not converge with increasing resolution. This is expected, since saturation requires the viscous scale to be resolved, while the Reynolds number in primordial gas clouds is of order $10^5$. Using cosmological initial conditions and a detailed chemical model, \citet{turk12} confirmed the results of \citet{sur10}, but found that $64$ instead of $32$ cells per Jeans length are necessary (see Figure~\ref{fig_mag}).

The turbulent dynamo rapidly amplifies the magnetic field to a level where it becomes dynamically important. In an accretion disk, a strong field can trigger the magneto-rotational instability, or lead to the formation of a hydromagnetic jet that removes material along the poles of the disk and aids the transport of angular momentum \citep{ms04, tb04, sl06}. The idealized, three-dimensional simulations of \citet{machida06} demonstrated that a jet indeed forms, and is capable of removing a significant amount of mass from the disk. Later simulations showed that strong magnetic fields may also suppress fragmentation \citep{mmi08, md13}. \citet{peters14} included sink particles and used a polytropic equation of state to model the thermal evolution of the gas. They found that dynamically important magnetic fields delay the onset of fragmentation by an order of magnitude compared to the case where no magnetic field is present.

Irrespective of the strength of the seed field, it appears that the turbulent dynamo amplifies the magnetic field rapidly enough that it becomes dynamically important already during the initial collapse phase. Simple one-dimensional calculations have shown that the effects of ambipolar diffusion are negligible at this point \citep[e.g.][]{schober12}. The magnetic field is expected to have the largest effect following the formation of the circumstellar disk. Magnetic braking and the launch of a hydromagnetic jet increase the rate of angular momentum transport through the disk and affects its ability to fragment. Exactly how effective these processes are depends on the initial strength of the magnetic field and how well the turbulent dynamo is resolved. In the near future, it may become possible to also include dissipative processes such as ambipolar diffusion in multi-dimensional simulations, which decrease the strength of the magnetic field.

\subsection{Cosmic rays}

A background of cosmic rays at high redshift is primarily produced by SNe, and may affect the formation of the first stars. Due to their long mean free paths, they can affect the chemical evolution of primordial gas on cosmological scales. \citet{sv04} and \citet{vs06} found that cosmic rays pre-ionize the gas, which in turn facilitates the formation of H$_2$ and HD. This may decrease the minimum halo mass required to cools the gas by an order of magnitude at $z\sim 20$. Next to the enhanced cooling, cosmic rays also directly heat the gas. In the calculations of \citet{sb07}, the additional cooling provided by the enhanced H$_2$ and HD formation rate dominates over the heating, allowing the gas in a minihalo to cool to the temperature of the CMB. As discussed in Section~\ref{sec_hd}, this may change the characteristic mass of the star. Similar results were found in \citet{jce07}, where a larger range of parameters was explored. \citet{nio14} investigated the combined effects of cosmic rays and LW radiation on atomic cooling halos. Since the cosmic rays enhanced the formation of H$_2$, a larger LW flux was necessary to suppress cooling and star formation prior to the onset of Ly-$\alpha$ cooling.

\subsection{Streaming velocities}

An important effect that has been neglected in most studies of primordial star formation is the relative velocity between the DM and gas sourced by baryon acoustic oscillations before recombination \citep{th10}. This relative velocity has also been termed the `streaming velocity'. At $z\gtrsim 1000$, sound waves propagate through the IGM and affect the DM and gas differently. While the gas pressure decelerates the gas, the DM only interacts gravitationally and maintains its velocity. The magnitude of the relative velocity is related to the effective sound speed of the gas, which is of the order of the speed of light before recombination due to Compton scattering. After recombination, the effective sound speed drops to that of a gas with a temperature of $\sim 10^4\,{\rm K}$, such that the streaming velocity becomes supersonic. A $1\sigma$-peak at $z\simeq 1000$ has a relative velocity of $v_{\rm rel}\simeq 30\,{\rm km}\,{\rm s}^{-1}$ and a Mach number of $\mathcal{M}\simeq 5$. Following recombination, the relative velocity decreases $\propto a^{-1}$, such that at $z\gtrsim 20$ the streaming velocity is close to $1\,{\rm km}\,{\rm s}^{-1}$. This is comparable to the virial velocity of minihalos, where the effect is expected to be strongest.

\begin{figure}
\centering
\includegraphics[width=11cm]{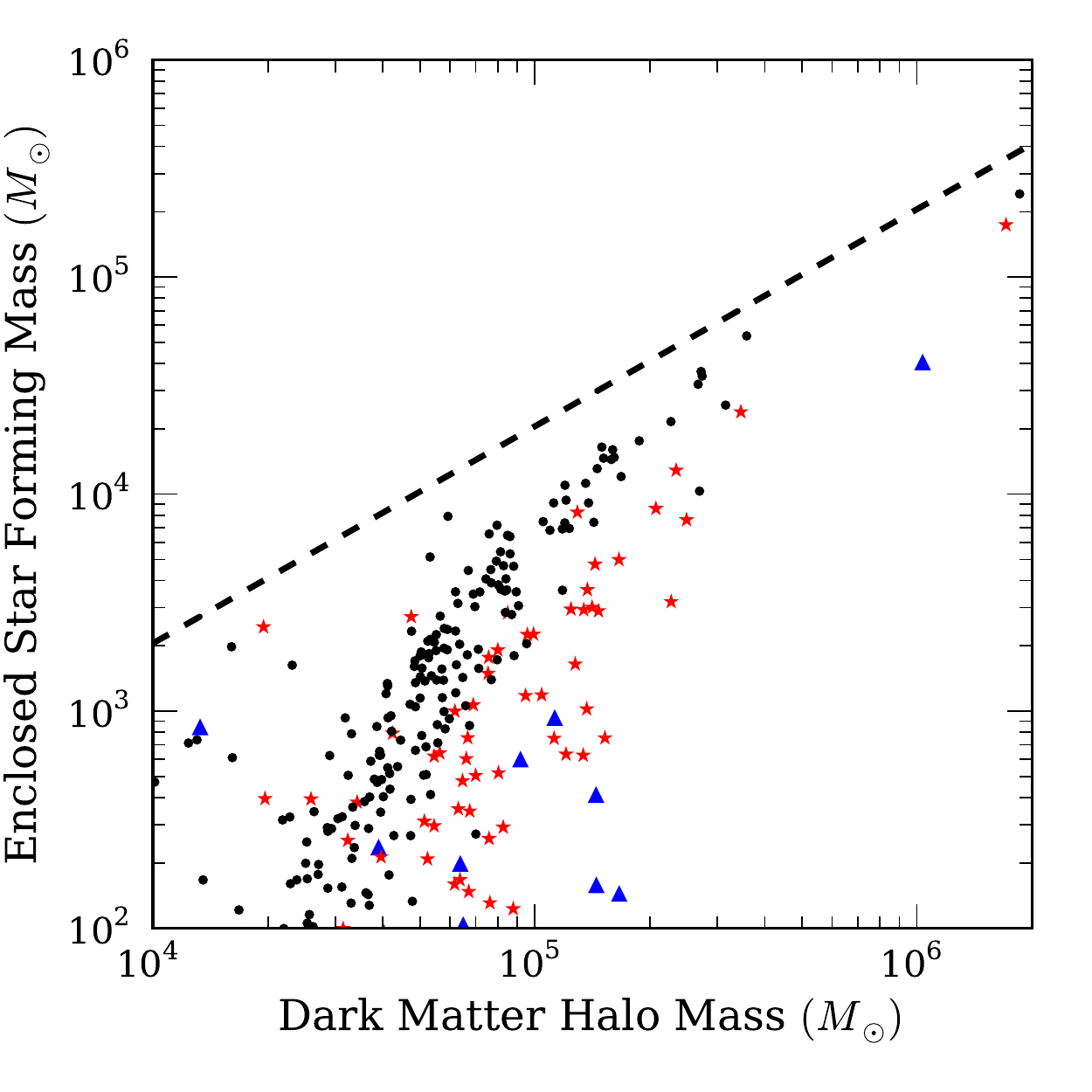}
\caption{Baryonic mass available for star formation in minihalos in a cosmological simulation that includes streaming velocities. The circles, stars, and triangles denote the outcome for Mach numbers $0$, $1.9$, and $3.8$ at $z=20$, respectively, and the black dashed line shows the gas mass assuming a cosmological baryon to dark matter ratio. As the streaming velocity is increased, the halos retain less gas, the central gas density is reduced, and less gas is able to cool and form stars. The degree to which star formation is suppressed increases with decreasing halo mass. Since the streaming velocities are sourced by acoustic oscillations before recombination, this leads to a modulation of Population~III star formation on $\simeq 10-100\,{\rm Mpc}$ scales. Adapted from \citet{om12}.}
\label{fig_stream}
\end{figure}

The streaming velocity damps density perturbations on the acoustic oscillation scale at recombination, i.e. in the range $\simeq 5-100\,{\rm Mpc}$. However, this effect is relatively small and only leads to a $\sim 10\%$ suppression of the number density of minihalos \citep{nyg12}. The effect on the virialization of the gas in minihalos is more pronounced. Numerical simulations included streaming velocities by either employing a fixed velocity offset between the DM and gas \citep{greif11b, mkc11, sbl11, nyg13, rst13, lns14}, or by self-consistently evolving the linear equations in the presence of streaming velocities \citep{om12}. They found that streaming velocities reduce the baryon overdensity in minihalos and possibly delay the onset of cooling (see Figure~\ref{fig_stream}). In addition, the center of the gas cloud may be shifted with respect to the center of the DM halo by a separation that is comparable to the virial radius. The moving DM halo induces a bow shock in the gas, which decelerates the halo via dynamical friction. Some of these effects were also found in more simplified semi-analytic calculations \citep{th10, tbh11, fialkov12}. The resulting modulation of star formation on large scales may also have a substantial effect on the 21-cm signal \citep{mo12, visbal12, fialkov12, fialkov13, fialkov14a}. \citet{tlh13} investigated the influence of streaming velocities on the formation and growth of stellar seed black holes (BHs), finding that they have only a minor effect on the abundance of BHs at late times.

\subsection{Dark matter annihilation}

Despite the fact that the cosmological mass density of DM is well constrained, its nature remains unknown. The most popular model invokes a WIMP, such as the lightest particle predicted by supersymmetry, which has a mass of $\sim 100\,{\rm GeV}$. These are expected to have a very small rate coefficient for self-annihilation of $\left<\sigma v\right>\simeq 3\times 10^{-26}\,{\rm cm}^3\,{\rm s}^{-1}$. However, this may be high enough for DM annihilations to have an effect on the DM and gas. Following a complex chain of reactions, the end products of DM annihilations are electron-positron pairs, neutrinos, and gamma rays, which can ionize and heat the gas. The average DM density is too low for this to have a significant effect on the IGM, but within halos the density-squared dependence of the annihilation rate may render DM annihilations important. In particular, minihalos are expected to have significantly higher central DM densities than halos in the present-day Universe. Their average density is significantly higher, since the virial density scales with the cube of the redshift. They are also more centrally concentrated, since the concentration parameter increases with decreasing halo mass \citep{nfw97}. Finally, their formation is expected to be nearly monolithic, since the variance of matter fluctuations is nearly constant towards the low-mass end. Following the collapse of the DM alongside the gas via `adiabatic contraction' \citep{young80, blumenthal86, gnedin04b}, DM annihilations begin to affect the gas.

\citet{sfg08} and \citet{freese09} used a simple dynamical model to show that in the standard WIMP scenario the DM heating rate matches the H$_2$ line cooling rate at a density of $n_{\rm H}\simeq 10^{13}\,{\rm cm}^{-3}$. They conclude that the collapse stalls at this point and a `dark star' powered by DM annihilations forms \citep[see also][]{nto09}. Stellar structure calculations that included DM-baryon scattering indicate that these stars are much larger than normal Population~III stars, have lower surface temperatures, and are more luminous \citep{freese08, iocco08b, iocco08c, taoso08, yia08, spolyar09, huy11, sg11}. Due to their low surface temperatures, they would appear dark at optical wavelengths, and radiative feedback would not be able to impede the accretion flow, allowing them to grow to a final mass of $\sim 10^6\,{\rm M}_\odot$. With a luminosity of up to $10^{10}\,{\rm L}_\odot$ in the infrared, they may be observable by the James Webb Space Telescope \citep[JWST;][]{freese10, zackrisson10a, zackrisson10b, ilie12}, next to other unique signatures \citep{sbk09b, maurer12, sandick12}.

One of the main problems of the dark star formation scenario is the assumption that the collapse of the gas stalls once the DM heating rate becomes comparable to the cooling rate. \citet{ripamonti10} showed that this is not the case in the standard WIMP scenario. Instead, the temperature increases only marginally until the cooling rate exceeds the DM heating rate, and the collapse continues nearly unhindered. Another problem concerns the inherent assumption of spherical symmetry. If the DM remains aligned with the gas, the annihilation rate is maximized. However, previous studies have shown that primordial gas clouds are permeated by transonic turbulence, develop a disk, and fragment in the later stages of the collapse \cite[e.g.][]{tao09, clark11a, greif12}. The influence of these perturbations on the DM profile was investigated by \citet{stacy12}. They found that they reduced the annihilation luminosity to a degree where it no longer had a substantial effect on the gas. While \citet{smith12b} found that the circumstellar disk was stabilized by annihilation heating, \citet{stacy14} attributed this to the use of a spherically symmetric DM density profile. When they allowed the DM potential to vary, they recovered the results of \citet{stacy12}.

In summary, while DM annihilation heating may affect the thermodynamic evolution of the gas at moderate densities by ionizing the gas and facilitating the formation of H$_2$ \citep[e.g.][]{ripamonti10}, it appears unlikely that the high-density evolution is changed to a degree at which the formation of dark stars instead of `normal' Population~III stars is favored.

\subsection{Alternative cosmologies}

Next to the standard $\Lambda$CDM paradigm, a number of alternative cosmologies may be viable. These could alter the formation of the first stars by changing the matter fluctuation power spectrum on small scales. In common gravitino warm dark matter (WDM) models, the particle mass is assumed to be $\gtrsim 1\,{\rm keV}$, since lower values are ruled out observationally \citep[e.g.][]{narayanan00}. \citet{yoshida03a} investigated the effects of a WDM particle with $m_{\rm WDM}=10\,{\rm keV}$ on the formation of the first stars. In this case, the matter power spectrum has an exponential cut-off at $\simeq 0.05\,{\rm Mpc}$, which corresponds to a mass of $\simeq 5\times 10^6\,{\rm M}_\odot$. As a result, they found that star formation in minihalos is nearly entirely suppressed, and must await the virialization of larger halos. \citet{on06} used somewhat more detailed simulations to investigate the influence of WDM particles with masses in the range $10\,{\rm keV}\gtrsim m_{\rm WDM}\gtrsim 50\,{\rm keV}$. At the lower mass end, they find a similar increase in the halo mass required for gas collapse as \citet{yoshida03a}. In addition, they found a delay in the onset of runaway cooling. However, once the gas collapsed to a density of $n_{\rm H}\simeq 10^5\,{\rm cm}^{-3}$, the thermodynamic evolution of the cloud became nearly indistinguishable from that in the CDM paradigm. \citet{gt07} used a WDM particle mass of $3\,{\rm keV}$ and found that the gas collapsed along a filament prior to the onset of runaway cooling \citep[see also][]{nu99, nu01, nu02, bt12, bt13}. Since the particle noise exceeds the WDM fluctuation power on small scales, these simulations could not derive the resulting fragment mass. \citet{uv14} improved upon these aspects and found that Population~III star formation was substantially suppressed at high redshifts. 

WDM particles consisting of sterile neutrinos also suppress small-scale power, but have the additional effect that they decay into X-rays, which ionize the gas and enhance the abundance of H$_2$ molecules \citep{bk06, sbk07}. This may facilitate Population~III star formation in minihalos. The influence of a running spectral index on the number density of minihalos was investigated by \citet{sbl03} and \citet{yoshida03c}. In these models, the power-law index of the primordial power spectrum decreases with increasing wavenumber. For a plausible 'running' of the spectral index, these studies found that the number density of minihalos was suppressed by about two orders of magnitude at $z=20$. Another possible deviation from the standard CDM power spectrum may come from non-Gaussianities. However, for realistic values of the dimensionless coupling constant $f_{\rm NL}$, \citet{mi11} and \citet{mk12} found that the properties of minihalos change only by a few percent. In quintessence models, the equation-of-state parameter $w$ is as a function of redshift, and is considered to decrease from a value $w>-1$ at $z>0$ to $w=-1$ at $z=0$, which results in enhanced small-scale power at high redshifts. For quintessence models that do not violate other cosmological constraints, the number density of minihalos may increase by up to an order of magnitude at $z=20$ \citep{maio06}.

\section{From the first stars to the first galaxies} \label{sec_fg}

\subsection{Definition}

The first stars are unambiguously associated with the first DM halos in which primordial gas is able to collapse and cool. The `first galaxies', however, are more difficult to define \citep[see][]{by11}. Since the term `galaxy' refers to an association of stars in a gravitationally bound system, a minihalo hosting a binary stellar system may already be considered a first galaxy. This definition may also be favorable from an observational standpoint, since the stellar radiation from star-forming minihalos may be spectroscopically indistinguishable from that from more massive halos. An alternative definition is based on the transition induced by the onset of atomic hydrogen cooling in halos with virial temperatures $\gtrsim 10^4\,{\rm K}$. In these halos, the onset of cooling does not depend on the presence on molecular hydrogen. In addition, gas photoionized and heated by stellar radiation remains gravitationally bound to the halo, which is generally not the case in minihalos. The self-sustaining cycle of star formation and feedback that is associated with galaxies can therefore operate in these halos.

Since this review is primarily concerned with the theory of primordial star and galaxy formation, I will use the definition involving the threshold for atomic hydrogen cooling. In this sense, the terms `first galaxy' and `atomic cooling halo' both refer to halos with virial temperatures $\gtrsim 10^{4}\,{\rm K}$. The corresponding relation between virial mass and virial temperature may be obtained via equation~\ref{eq_tvir}:
\begin{equation}
M_{\rm vir}\simeq 5\times10^7\,{\rm M}_\odot\left(\frac{T_{\rm vir}}{10^4\,{\rm K}}\right)^{3/2}\left(\frac{1+z}{10}\right)^{-3/2}.
\end{equation}
The characteristic virial mass of an atomic cooling halo is therefore $10^7\lesssim M_{\rm vir}\lesssim 10^8\,{\rm M}_\odot$, with a typical formation redshift of $z\simeq 10$--$15$ for $2$--$3\sigma$ peaks.

\subsection{Turbulence}

The virialization of primordial gas in atomic cooling halos without prior star formation in minihalos has been investigated by \citet{wa07b} and \citet{greif08}. They found that the gas accreted onto the halo first shock-heats to the virial temperature, after which Ly-$\alpha$ cooling is activated and virial equilibrium is attained via turbulence. At higher densities, H$_2$ line cooling takes over and the turbulence becomes supersonic with Mach numbers $M\simeq 5$. This is an important difference to minihalos, where the gas is at most mildly supersonic. \citet{prieto11} showed that this turbulence leads to the formation of a number of gravitationally bound clumps within the halo. \citet{wa07b} and \citet{greif08} also found that two distinct modes of accretion exist. In the standard `hot accretion' mode, the gas accretes nearly radially onto the halo and shock-heats to the virial temperature near the virial radius \citep{bd03}. This is accompanied by a `cold accretion' mode, in which cold intergalactic gas accumulates in filaments before streaming onto the halo \citep{keres05, db06}. In the simulations of \citet{wa07b}, these streams reach down to about $25\%$ of the virial radius, while they penetrate the center of the halo in \citet{greif08}. In the latter study, the prominence of cold streams may have been overestimated due to the aggressive accretion of the gas by a massive BH at the center of the halo, and the artificial suppression of mixing \citep{nelson13, fernandez14}.

\subsection{Radiative feedback}

Analytic considerations as well as simulations have indicated that the first stars had masses $M_*\sim 100\,{\rm M}_\odot$, possibly with a large scatter around this value. For the purpose of stellar evolution, I therefore only consider massive Population~III stars, even though calculations for their low-mass counterparts exist \citep{chieffi01, gs01, sll02, gil-pons05, ggg07, sfi07, lawlor08, lst08, mocak10}. Massive Population~III stars have low opacities due to the absence of metals, and ignite nuclear burning at very high temperatures as a result of inefficient proton-proton and CNO burning \citep[e.g.][]{efo83, bac84, marigo01}. They are therefore expected to be smaller and hotter than Population I/II stars of the same mass. The spectral shape of the radiation emitted by Population~III stars on the main sequence may be derived by combining stellar structure calculations with detailed LTE and non-LTE model atmospheres \citep{cojazzi00, ts00, bkl01, schaerer02, schaerer03}. These studies concluded that massive Population~III stars radiate approximately as blackbodies with an effective temperature of $\simeq 10^5\,{\rm K}$, and produce up to an order of magnitude more UV photons per stellar baryon than normal stars. Depending on mass, they emit most of their radiation in the LW bands in the range of $\simeq 11.2$--$13.6\,{\rm eV}$, or above the H\,{\sc i}, He\,{\sc i}, or He\,{\sc ii} ionizing thresholds of approximately $13.6$, $24.6$, and $54.4\,{\rm eV}$, respectively. They are also strong emitter of X-ray radiation. Although important for the 21-cm signal and the opacity of the IGM, I will here not discuss Ly-$\alpha$ radiation \citep[for a review, see][]{dijkstra14}.

\subsection{Photodissociating radiation}

One possible way to prevent the formation of H$_2$ is to photodetach H$^-$, which is one of the main reaction partners in forming H$_2$. However, since the reaction rate for associative detachment is so large, this process is usually not important \citep[but see][]{cks07, glover07, wh12}. Furthermore, the direct dissociation of H$_2$ via radiative excitation to the vibrational continuum is highly forbidden. Most of the H$_2$ is therefore dissociated by the two-step Solomon process \citep{fsd66, sw67}. Radiation in the LW bands excites a higher electronic state of H$_2$, which is followed by decay to the vibrational continuum in approximately $15\%$ of all cases. Detailed radiative transfer calculations have shown that the optically thin dissociation rate can be written as $k_{\rm diss}\simeq 10^8F_{\rm LW}\,{\rm s}^{-1}$, where $F_{\rm LW}$ is the average flux density in the LW bands in ${\rm erg}\,{\rm s}^{-1}\,{\rm cm}^{-2}\,{\rm Hz}^{-1}$ \citep{db96, abel97}.

Initial studies showed that the LW radiation from a single massive Population~III star is sufficient to prevent further cooling and star formation in the halo in which the star formed \citep{on99, nt00, gb01}. Subsequent work concentrated on the effects of LW radiation on cosmological scales, finding that the H$_2$ in the IGM with a fractional abundance of only $y_{\rm H_2}\sim 10^{-6}$ is quickly dissociated. The optically thin IGM then becomes permeated by a global LW background, which may suppress star formation in halos with virial temperatures below $\simeq 10^4\,{\rm K}$ well before reionization \citep{hrl97, kbs97, cfa00, har00, kitayama01}. However, only a modest intensity of the order of $J_{\rm LW}=10^{-23}\,{\rm erg}\,{\rm s}^{-1}\,{\rm cm}^{-2}\,{\rm Hz}^{-1}\,{\rm sr}^{-1}$ is built up by $z\simeq 20$, which is not sufficient to quench star formation in minihalos \citep{su00, kitayama01, rgs02b, yoshida03a, gb06, mst06, jgb08, ts09}. Instead, the minimum virial mass required for efficient H$_2$ cooling increases relatively slowly \citep{mba01, yoshida03a, mbh06, jgb07, wa07a, on08, latif14b, visbal14}. Unless the LW flux is extremely high, H$_2$ cooling may even become important in atomic cooling halos \citep{sk00, omukai01, oy03, oh02, whb11, safranek-shrader12}.

Even though the LW flux is nearly uniform when averaged over cosmological distances, the flux seen by individual halos may fluctuate by many orders of magnitude due to their large formation bias \citep{dijkstra08, ahn09, hf12}. This may allow the flux to become high enough to completely suppress the formation of molecules in atomic cooling halos, possibly resulting in the formation of direct collapse BHs (see Section~\ref{sec_dc}). The combination of photodissociating and photoionizing radiation may enhance the formation of H$_2$ in a thin shell ahead of an H\,{\sc ii} region, which may trigger cooling and collapse in nearby halos \citep{hrl96, rgs01, kitayama04, as07, su06, susa07, suh09, whalen08a, whm10}. Since this positive feedback effect requires fine-tuning of various relevant parameters, such as the distance of the star and the central density of the halo, this scenario is likely not of cosmological significance. Another argument against a positive feedback effect was provided by \citet{glover07}, who showed that recombination radiation from the H\,{\sc ii} region may suppress H$_2$ formation in the thin shell ahead of the H\,{\sc ii} region due to the dissociation of H$^-$ and H$_2^+$.

\subsection{Ionizing Radiation}

Photoionizing radiation from Population~III stars has a strong effect on the gas in the halos in which they form. The first calculations that investigated the propagation of ionizing radiation from Population~III stars in minihalos used a simple dynamical model, but solved the radiative transfer accurately \citep{kitayama04, wan04}. They found that the ionizing radiation is initially trapped well within the halo by a D-type ionization front, which drives a hydrodynamic shock with a speed of $\simeq 30\,{\rm km}\,{\rm s}^{-1}$ that begins to blow out the gas. After $\simeq 10^5\,{\rm yr}$, the shock has nearly evacuated the halo, and the ionization front becomes R-type and propagates into the IGM. The resulting density and velocity profiles are comparable to the self-similar solutions for the champagne flows discussed in \citet{shu02}. Later three-dimensional simulations qualitatively confirmed these results, and found that $\simeq 100\,{\rm M}_\odot$ Population~III stars create H\,{\sc ii} regions with sizes of a few kpc, and photon escape fractions that approach unity \citep{oshea05, abs06, awb07}. In the simulation of \citet{greif09b}, the H\,{\sc ii} region breaks out anisotropically due to the presence of a disk. \citet{yoshida07} included helium-ionizing radiation and found that a significant He\,{\sc ii} region with a temperature in excess of $3\times 10^4\,{\rm K}$ develops.

The relic H\,{\sc ii} region left behind after the star fades away cools faster than it recombines. The electron fraction therefore remains high even after the gas has cooled to below $\simeq 10^4\,{\rm K}$. The elevated electron fraction facilitates the formation of H$_2$ to a level of $y_{\rm H_2}\simeq 10^{-3}$, which allows the gas to cool to temperatures where chemical fractionation occurs. The enhanced HD abundance then facilitates cooling to the temperature of the CMB \citep{jb07, jgb07, yoshida07, yoh07}. As discussed in Section~\ref{sec_hd}, metal-free stars that form in relic H\,{\sc ii} region gas may have a lower characteristic mass than stars that form in minihalos unaffected by radiation. The time scale for relic H\,{\sc ii} region gas to re-collapse is of order the Hubble time, such that continued star formation in a minihalo must await the virialization of larger halos.

The influence of ionizing radiation on neighboring halos has been investigated as well. The sign of the feedback depends on various parameters, such as the state of the collapse and the distance to the source. If a halo is close to an ionizing source or has not yet collapsed to high densities, the halo may be photoevaporated, while in other cases the halo may survive and continue to collapse \citep{ki00, kitayama00, kitayama01, sk00, su00, su04, su06, suh09, oshea05, mbh06, mbh09, whalen08a, whm10, hus09}. \citet{wn08} and \citet{vasiliev12a} also showed that shadow and thin-shell instabilities may develop in the ionization fronts. Once a pervasive UV background has been established, star formation in minihalos will ultimately be shut down.

The UV radiation of massive Population~III stars begins the process of reionization \citep[e.g.][]{go97, hl97, ciardi00, ciardi01, wl03a, sir04, sokasian04, wc07, aft12, ahn12}. While minihalos likely did not contribute significantly to the total ionizing photon budget at high redshifts, the first galaxies are expected to have been much more important \citep{bl01, cen03a, cen03b, cfw03, oh03, ro04a, fl05, gb06, hb06, johnson09, ts09, vb11}. Simulations that model the propagation of ionizing radiation from individual Population~III star clusters typically find very high escape fractions that are of the order of unity \citep{wa08a, wa08b, wc09, wise14, rs10, pkd13}. The ionizing radiation may even suppress star formation in atomic cooling halos \citep{bl99, tw96, gnedin00, ki00, kitayama01, dijkstra04}. Another interesting effect stems from minihalos that are not massive enough to host star formation. These become increasingly common as reionization proceeds, and may act as sinks of ionizing radiation \citep{ham01, bl02a, isr05, iss05, ciardi06}.

\subsection{X-rays} \label{sec_xrays}

Although the direct emission of X-rays from Population~III stars is likely cosmologically unimportant \citep[e.g.][]{vb11}, they may indirectly contribute to the build-up a pervasive X-ray background. One of these sources is the accretion of gas onto the compact remnants of Population~III stars. These `miniquasars' may pre-heat and ionize the IGM due to the long mean free path of photons with energies $\ge 1\,{\rm keV}$ \citep{hl99, vgs01, gb03, mba03, madau04b, ro04b, rog05, shf05, tph12}. They may also increase the intergalactic H$_2$ abundance by more than an order of magnitude and reduce the clumping of the IGM \citep[e.g.][]{km05b}. Similar to the evolution of relic H\,{\sc ii} regions, the enhanced H$_2$ abundance may lead to chemical fractionation of HD \citep{hummel14}. On smaller scales, the radiation pressure and heating from the quasar reduces the surrounding density and impedes the accretion flow onto the BH \citep{jb06, awa09, milosavljevic09, mcb09, pr11, pr12, aykutalp13}. It is therefore unlikely that miniquasars will grow fast enough to explain the presence of super-massive BHs at $z\gtrsim 6$ \citep{fck06}. However, their radiation may delay star formation in neighboring minihalos until the atomic cooling threshold is surpassed \citep{awa09, jeon12}.

Another source of X-rays is Roche lobe overflow in a binary system. For massive Population~III stars, the collapse of the star into a BH is more likely than in the Population~I/II case \citep[e.g.][]{heger03}. In addition, a number of studies have shown that a significant fraction of Population~III stars in minihalos may have formed in binaries \citep{smu04, machida08a, machida08b, machida09, tao09, sgb10, gsb13, sb13}. Although the spectrum is different, the effects of the radiation from X-ray binaries are similar to those of miniquasars \citep{power09, jeon12, power13, jeon14a, xu14}. X-rays may also be produced by Population~III SNe as a result of thermal bremsstrahlung or inverse Compton scattering of relativistic electrons off the CMB \citep{oh01a, gb03}.

\subsection{Final fates of Population~III stars} \label{sec_sn}

\begin{figure}
\centering
\includegraphics[width=12cm]{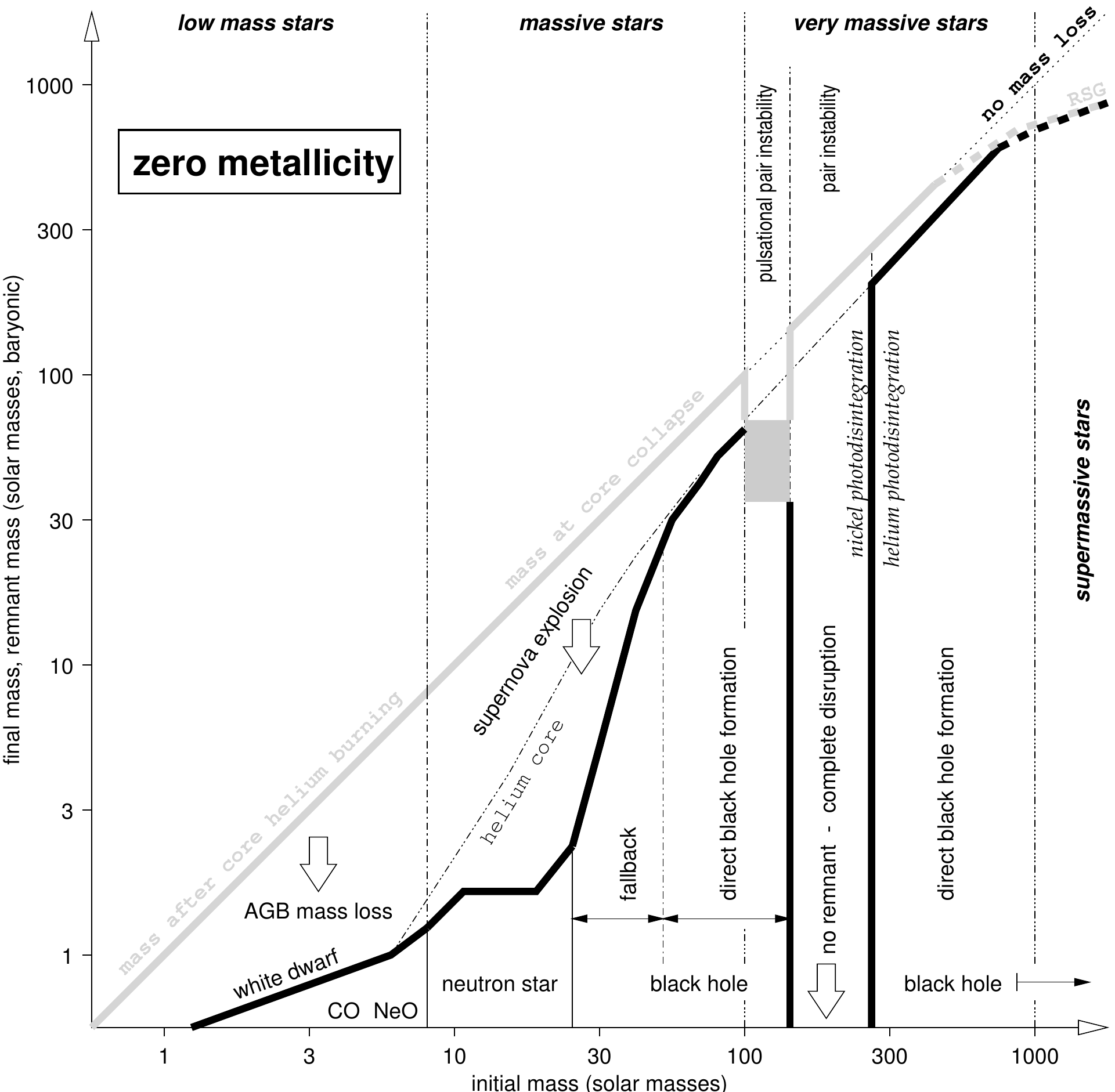}
\caption{Final fates of Population~III stars in the absence of stellar rotation. Above $\simeq 30\,{\rm M}_\odot$, a fraction of the star may collapse to a BH, while at higher masses it collapses directly to a BH or explodes as a pair-instability SN. The chemical yields of the ejecta depend sensitively on the fate of the star. The mass limits change if rotation is taken into account. Adapted from \citet{hw02}.}
\label{fig_sn}
\end{figure}

Massive Population~III stars burn their nuclear fuel very quickly and live only a few million years \citep[e.g.][]{bac84}. Models of non-rotating Population~III stars have shown that in the mass range $\simeq 140-260\,{\rm M}_\odot$, a so-called pair-instability SN disrupts the entire star \citep{fwh01, hw02, heger03, jw11, baranov13, chen14a, chen14c}. In this case, the center of the star loses pressure support due to the creation of electron-positron pairs. This leads to explosive nucleosynthesis, which produces a metal yield of $\simeq 50\%$ and a kinetic explosion energy of up to $10^{53}\,{\rm ergs}$. In the range $\simeq 100$--$140\,{\rm M}_\odot$, pulsational instabilities drive episodic outbursts, while in the range $\simeq 40$--$100\,{\rm M}_\odot$ the entire star collapse to a BH. Below $\simeq 40\,{\rm M}_\odot$, a SN with $\sim 10^{51}\,{\rm ergs}$ partially disrupts the star. Depending on mass, a fraction of the star collapses into a BH, which leads to an elemental segregation of the nucleosynthetic products \citep{cl02, cl04, un02, un03, un05, iwamoto05, tun07, zwh08, jwh09, hw10, jaw10, lc12}. For masses $\gtrsim 260\,{\rm M}_\odot$, the entire star collapses directly to a BH without any significant SN explosion \citep[but see][]{ohkubo06, iho13}. Finally, super-massive stars with $\simeq 10^5$--$10^6\,{\rm M}_\odot$ may form in atomic cooling halos in which previous star formation was suppressed. A general relativistic instability develops and a fraction of the star may collapse into a BH with $\simeq 10^4$--$10^5\,{\rm M}_\odot$ \citep{heger03, bvr06, bra08, begelman10, vb10, mjm12, hoy12, volonteri12, iho13, hosokawa13, schleicher13, chen14b}. Recent studies have found that a super-massive star may also trigger an extremely energetic SN explosion with up to $10^{55}\,{\rm ergs}$ of kinetic energy \citep{whalen13d, whalen13e, whalen13g}. The various fates of Population~III stars are illustrated in Figure~\ref{fig_sn}.

Most of the above studies have neglected the effects of rotation. Models including rotation show that metals may be mixed between nuclear burning layers and even to the surface of the star. This may have an effect on the evolution of the stars, the degree to which their elements are mixed \citep{mm02a, mm02b, mem06, hws05, chiappini06, hirschi07, chiappini08, ekstroem08, tuy14}, and their final fates \citep{suwa07a, joggerst10, cw12a, cw12b, ydl12, cwc13}. For example, \citet{cw12a} found that strongly rotating Population~III stars may already explode as pair-instability SNe if their mass exceeds $\simeq 65\,{\rm M}_\odot$. If a significant amount of rotation persists until the collapse phase, the SN explosion may be accompanied by a gamma-ray burst \citep{fwh01, mn03, yl05, yln06, tominaga07, tominaga09, kb10, si11, nsi12, smidt14}.

\subsection{Mechanical Feedback}

The kinetic energy released by the SN of a massive Population~III star can have a substantial effect on the halo in which the progenitor star formed. \citet{ky05} and \citet{whalen08b} used one-dimensional calculations to investigate the evolution of SN remnants in minihalos with various explosion energies. They found that more conventional core-collapse SNe fail to remove the gas from more massive halos, while pair-instability SNe are able to completely evacuate even the most massive halos. The expansion of the remnant also depends sensitively on the presence of an H\,{\sc ii} region, which reduces the central density prior to the explosion. Numerical simulations showed that the remnant of an energetic pair-instability SN expands to a maximum radius of a few kpc, which is comparable to the size of the H\,{\sc ii} region created by the progenitor star \citep{byh03, greif07, wa08b, sbs14}. In nearly all cases, the expansion can be divided into three distinct phases. In the free expansion phase, the momentum of the swept-up gas is negligible compared to that of the remnant. Once the inertia of the ambient medium becomes important, the remnant enters the energy-conserving Sedov-Taylor phase \citep{taylor50, sedov59}. Finally, radiative losses facilitate the transition to the momentum-conserving snowplow phase, where the expansion is driven solely by the inertia of remnant. One of the most important coolants in the final phase is inverse Compton scattering. This process is particularly important in the high-redshift IGM, due to the strong dependence of the cooling rate on the temperature of the CMB.

The chemical and thermal evolution of the gas in the SN remnant is similar to that in relic H\,{\sc ii} regions. The elevated electron abundance facilitates the formation of H$_2$ and HD, which may trigger secondary Population~III star formation once the gas recollapses. For highly energetic explosions, the time required for the gas to recollapse is of the order of the Hubble time \citep{greif10}, while for less energetic explosions the collapse time is $\simeq 10\,{\rm Myr}$ \citep{ritter12}. Idealized simulations have also found that fragmentation in the dense shell swept up by the SN remnant may occur \citep{sfs04, machida05, vvs08, nho09, cyk13}. The influence of the remnant on neighboring halos depends primarily on the distance of the halo from the progenitor star and their density \citep{greif07, ss09, whm10}. If they are close enough and have not yet collapsed, star formation will be delayed or suppressed, while in rare cases the shock wave may compress the halo and trigger collapse. Recent studies also investigated the explosion of super-massive stars with up to $10^5\,{\rm M}_\odot$ in atomic cooling halos that remained metal-free \citep{johnson13a, whalen13d, whalen13e}. These studies employed various methods to evolve the SN remnant through the distinct stages, and investigated their impact on the progenitor halo. With a kinetic energy of up to $10^{55}\,{\rm ergs}$, these SNe were able to completely disrupt their host halos.

\subsection{Chemical Enrichment}

In addition to their mechanical feedback, SN explosions from the first stars enrich the IGM with metals \citep{mfr01, std01, mfm02, sfm02, mbh03, ssf03, wv03, ro04a, scannapieco05b}. The chemical yield depends sensitively on the type of the SN. For example, a pair-instability SN mainly produces elements with an even nuclear charge, and almost no neutron-capture elements, while more conventional core-collapse SNe display a characteristic enhancement of $\alpha$-elements. The enrichment pattern of the IGM also depends on the SN explosion energy and on how many SN remnants overlap prior to the re-collapse of the enriched gas. \citet{greif08} found that of order $10$ star-forming minihalos merge to form a first galaxy, which implies that a similar number of SN ejecta mix with each other prior to second-generation star formation. Based on numerical simulations that modeled the formation and explosion of isolated Population~III stars in minihalos, a number of studies found that the gas is quickly enriched to a metallicity of $Z\sim 10^{-3}\,{\rm Z}_\odot$ \citep{wa08b, greif10, ritter12, wise12a, vasiliev12b, chen14d}. In a recent study, \citet{ritter14} employed tracer particles at very high resolution to follow the evolution of metal-enriched gas. They found that mixing in the IGM is suppressed due to the long eddy turnover time, while the turbulence associated with the virialization of the underlying DM halo facilitates complete mixing within the halo. In addition, the differing yields of the various SN mass shells are reflected in the enrichment pattern of the re-collapsing gas, which prevents a one-to-one mapping between the nucleosynthetic yield of the SN and the stars that form from its remnants.

\begin{figure}
\centering
\includegraphics[width=13cm]{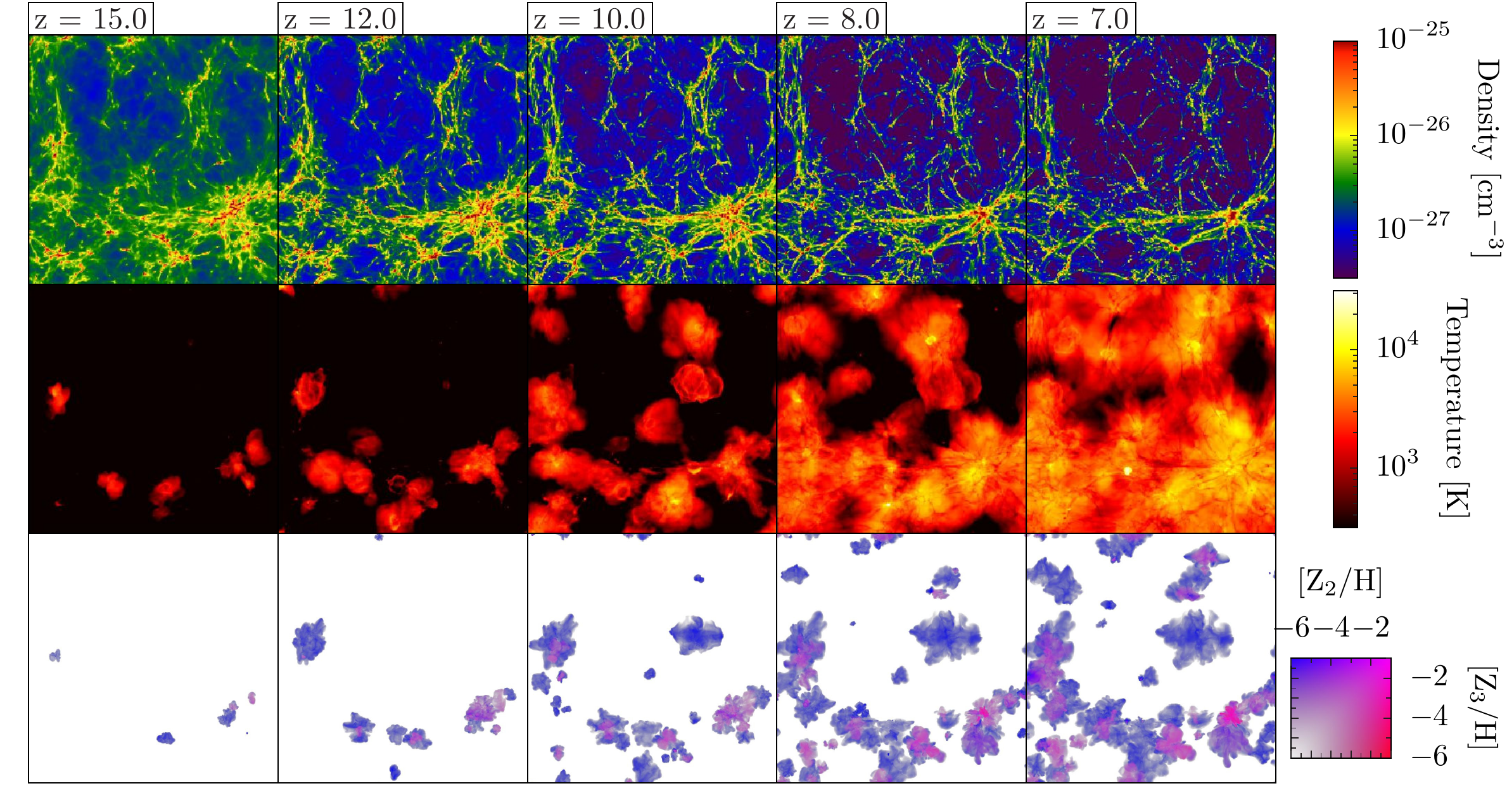}
\caption{Three-dimensional simulation that models the transition from Population~III to Population~I/II star formation on cosmological scales. The physical model includes ionizing radiation from various stellar populations, and the mechanical energy input and chemical enrichment from SNe. The columns denote the redshift, and the rows the mass density, temperature, and metallicity originating from Population~III and Population~I/II stars, respectively. The box size is $1\,{\rm Mpc}$ (comoving). Star formation takes place in halos with masses between $\simeq 10^6$ and $\simeq 10^9\,{\rm M}_\odot$, and gradually enriches the IGM with metals. Adapted from \citet{wise12a}.}
\label{fig_met}
\end{figure}

The metal-enriched gas tends to re-collapse on a time scale of $10$--$100\,{\rm Myr}$ as the underlying atomic cooling halo virializes, such that the transition to Population~I/II star formation occurs very rapidly \citep{jeon14b}. In fact, studies that investigated metal enrichment on cosmological scales found that the global star formation rate is dominated by normal stars already at $z\sim 20$ \citep{gb06, ts09, crosby13, jdk13, muratov13b, xwn13}. However, due to the high spatial bias of minihalos, the enrichment of the IGM proceeds very anisotropically, with pockets of Population~III star formation surviving to very low redshifts \citep{nop04, tfs07, rgs08, maio10, maio11, muratov13a, muratov13b}. Figure~\ref{fig_met} shows the patchy enrichment of the IGM in the simulation of \citet{wise12a}. Earlier models assumed that the ejected metals promptly enriched the IGM and established a bedrock metallicity of the order of $10^{-5}\,{\rm Z}_\odot$ \citep{oh01b, schneider02, mbh03, vt03, fc04, ro04a, ybh04, mc05, gb06}. Despite the progress made, the degree to which the metals mix with primordial gas is not yet fully understood \citep{pss13}, and depends on the employed sub-grid model \citep{greif09a, ritter12, ritter14}.

\subsection{Critical Metallicity}

\begin{figure}
\centering
\includegraphics[width=7cm]{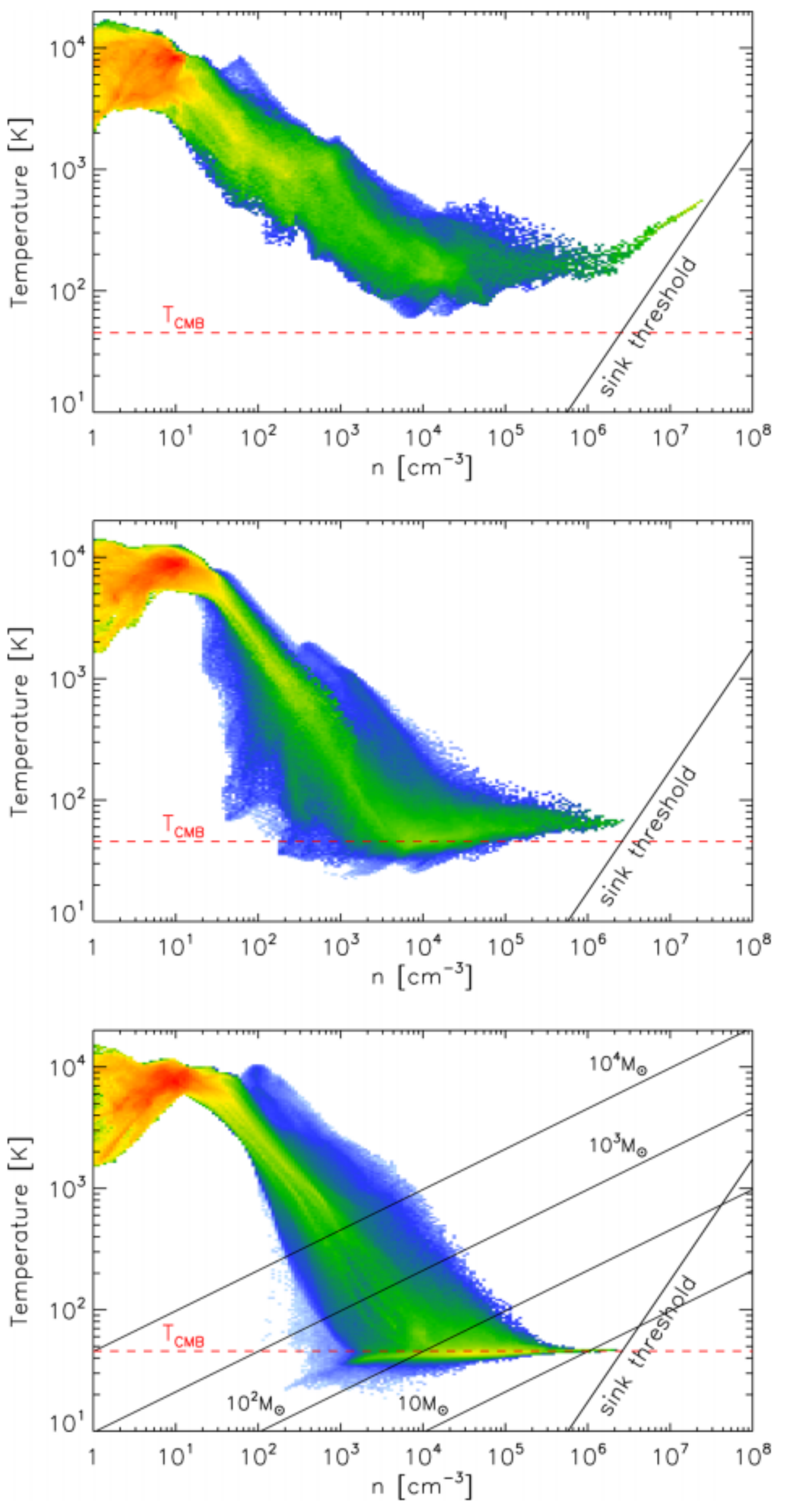}
\caption{Thermodynamic evolution of metal-enriched gas in an atomic cooling halo at $z\simeq 16$. The panels show the results for metallicities $Z=10^{-4}$, $10^{-3}$, and $10^{-2}\,{\rm Z_\odot}$, respectively (top to bottom). The gas mass per bin over the entire mass in the box is color-coded from blue (lowest) to red (highest). The dashed red lines show the temperature of the CMB, and the solid black lines on the right-hand side denote the threshold for sink particle formation. In the bottom panel, lines of constant Jeans mass are indicated as well. As the metallicity increases, metal fine-structure cooling allows the gas to cool to lower temperatures. This decreases the characteristic Jeans mass of the fragments, and facilitates the transition from Population~III to Population~I/II star formation. Adapted from \citet{smb14b}.}
\label{fig_sf}
\end{figure}

The nature of the transition from Population~III to Population~I/II star formation remains a matter of debate. \citet{bromm01} argued that the fine-structure cooling provided by carbon and oxygen allows the gas to cool to the temperature of the CMB once the gas metallicity exceeds $Z\sim 10^{-3}\,{\rm Z}_\odot$, which reduces the characteristic Jeans mass of the cloud and facilitates fragmentation \citep[see also][]{bl03a, ss06a, sbm10}. Later simulations included H$_2$ cooling and confirmed that metal line cooling reduces the fragment mass \citep[see Figure~\ref{fig_sf};][]{ss07, ssa08, smith09, smb14b}. Other studies argued that H$_2$ cooling is just as important as fine-structure cooling at the relevant densities and temperatures, and that there is no clear distinction in the resulting fragment masses \citep{jappsen07, jappsen09a, jappsen09b}. However, these studies were carried out at very high redshifts where the CMB temperature was close to the minimum temperature that may be reached via H$_2$ line cooling. Large differences in the cooling rates thus did not have a substantial effect on the thermodynamic evolution of the clouds. In addition, metal fine-structure cooling is relatively insensitive to the strength of the UV background, while molecules may be easily destroyed at the redshifts at which the first metal-enriched stars form. This effect was confirmed by \citet{bovino14b}, who found that halos with a metallicity greater than $\simeq 10^{-3}\,{\rm Z}_\odot$ cooled to the temperature of the CMB in the presence of a strong UV background that dissociated the H$_2$.

Another possible trigger for a transition to normal star formation is dust cooling. In this case, the critical metallicity is expected to be as low as $Z\sim 10^{-6}\,{\rm Z}_\odot$. At high redshifts, dust is thought to be produced in the SN remnants of Population~III stars \citep{tf01, nozawa03, sfs04, nkh06, nozawa07, gah11a, gah11b, nkn12, nozawa14}. A characteristic mass of $\sim 1\,{\rm M}_\odot$ may only be obtained for dust cooling, since the dip in the effective equation of state lies at much higher densities than for fine-structure cooling \citep{omukai00, schneider03, omukai05, schneider06, to08, ohy10, so10, schneider12, cny13, chiaki14}. \citet{to06} modeled dust cooling in three-dimensional simulations using idealized initial conditions and found that the central clump becomes elongated and fragments for $Z\gtrsim 10^{-6}\,{\rm Z}_\odot$. \citet{cgk08} started from more realistic initial conditions, employed sink particles, and used a tabulated equation of state to model the effects of dust cooling. They found that the number of fragments greatly increases for $Z\gtrsim 10^{-5}\,{\rm Z}_\odot$. \citet{dopcke11} and \citet{dopcke13} improved upon the simulations of \citet{cgk08} by explicitly modeling the dust temperature, and found that fragmentation occurs at all metallicities, but is much more prominent if dust cooling becomes effective. \citet{mso14} investigated the combined effects of metal-line cooling and grain-catalyzed molecule formation, finding that the gas temperature approaches the CMB temperature at densities that decrease as the metallicity is increased, due to metal fine-structure cooling and the formation of H$_2$ on dust grains \citep[see also][]{lss12}. They also found that the amount of fragmentation increases with increasing metallicity. In a simulation that started from cosmological conditions and included metal fine-structure cooling as well as dust cooling, \citet{smb14a} found that realistic turbulent velocities in the first galaxies enhance the density contrast well above that expected from monolithic collapse models, reducing the characteristic fragment mass to $\simeq 0.1\,{\rm M}_\odot$.

The above studies show that the transition from Population~III to Population~I/II star formation is not only governed by the critical metallicity. Among many other factors, the mass function of the first metal-enriched stars depends on the initial conditions, the temperature of the CMB, the radiation background \citep{as11}, the metallicity, the elemental composition of the metals, and the dust depletion factor. Despite this complexity, metal fine-structure cooling typically results in fragment masses of the order of $10\,{\rm M}_\odot$, while dust cooling can lead to the formation of sub-solar or solar-mass fragments, since it operates at significantly higher densities.

\subsection{Direct collapse black holes} \label{sec_dc}

If the local LW flux is high enough, the formation of H$_2$ in minihalos may be suppressed until the halo has become massive enough for Ly-$\alpha$ cooling to become important \citep{omukai01, bl03b, vr05, ss06b, ssg10, johnson13b}. The gas may then contract isothermally at a temperature of $\simeq 10^4\,{\rm K}$ to $n_{\rm H}\simeq 10^6\,{\rm cm}^{-3}$, when the Ly-$\alpha$ radiation becomes trapped \citep[e.g.][]{lzs11}. At this point, two-photon emission and H$^-$ continuum cooling take over and extend the near-isothermal collapse phase to $n_{\rm H}\simeq 10^{16}\,{\rm cm}^{-3}$ \citep{omukai01}. Star formation in the progenitor halos of the atomic cooling halo must be suppressed, since the gas would otherwise become enriched with metals and have a low-temperature coolant \citep{jgb08, osh08}.

\begin{sloppypar}

Initial calculations showed that the critical LW flux required to suppress H$_2$ cooling for a photospheric temperature of $10^5\,{\rm K}$ corresponding to Population~III stars is of the order of $10^3$ in units of $J_{21}$, where $J_{21}=10^{-21}\,{\rm erg}\,{\rm s}^{-1}\,{\rm cm}^{-2}\,{\rm Hz}^{-1}\,{\rm sr}^{-1}$ \citep{omukai01}. A flux of this level dissociates H$_2$ up to a density of $n_{\rm H}\simeq 10^4\,{\rm cm}^{-3}$, where the level populations of H$_2$ transition to LTE and H$_2$ cooling becomes comparatively inefficient (see Section~\ref{sec_h2}). \citet{whb11} found that the critical LW flux may be reduced by an order of magnitude if a more accurate self-shielding formula for H$_2$ is used. The inclusion of radiation from Population~I/II stars and the treatment of H$^-$ dissociation further reduces the critical flux \citep{wh12}. \citet{sbh10} find a value of only $J_{\rm 21,crit}\simeq 30-300$ for a photospheric temperature of $10^4\,{\rm K}$ corresponding to Population~I/II stars. On the other hand, \citet{latif14a} employ higher resolution, a more accurate H$_2$ self-shielding formula, and investigate halos that virialize at slightly higher redshifts. They find a critical flux of $J_{\rm 21,crit}\sim 10^3$. \citet{ak15} and \citet{soi14} use a realistic stellar spectrum, which differs significantly from a blackbody spectrum, and find that this increases the critical flux to $J_{\rm 21,crit}\sim 10^3$, while \citet{vs13} argue that magnetic fields and turbulence potentially reduce $J_{\rm 21,crit}$ by an order of magnitude. Recent work has shown that the critical flux from a realistic metal-enriched population at high redshifts has a radiation temperature closer to $10^5$ than $10^4\,{\rm K}$ \citep{soi14}. \citet{latif14c} accounted for this fact and found that hydrodynamical effects such as shock-heating and inhomogeneous collapse, which are only captured in realistic three-dimensional simulations, increase the critical LW flux to a few times $10^4$ instead of $10^3$. Finally, \citet{rjw14} have argued that the radiation from a point source instead of a uniform background may further increase the critical LW flux.

\end{sloppypar}

The global LW background at high redshifts is much too low to reach these values \citep[e.g.][]{gb06}. However, the local LW flux can be boosted by many orders of magnitude near star formation sites \citep{dijkstra08, agarwal12, dfm14, agarwal14b}. The timing of the incident LW flux with respect to the state of the collapse of the atomic cooling halo is an important factor \citep{vhb14b}. The expected number density of halos exposed to a super-critical flux is still highly uncertain. \citet{agarwal12} used $J_{\rm 21,crit}=30$ and found $\sim 10^{-3}$ direct collapse black holes per comoving Mpc at $z=10$, while \citet{dfm14} used $J_{\rm 21,crit}=300$ and included metal enrichment, finding a number density of $10^{-10}-10^{-5}\,{\rm Mpc}^{-3}$ (comoving) at $z=10$, respectively. These numbers seem more realistic in light of recent work showing that the critical flux is closer to $J_{\rm 21,crit}=10^4$ \citep[e.g.][]{latif14c, soi14, yue14}.

Other mechanisms for suppressing H$_2$ formation and cooling have been suggested as well. \citet{io12} proposed that cold accretion flows may penetrate deep into atomic cooling halos and shock-heat the gas at densities $n_{\rm H}\gtrsim 10^4\,{\rm cm}^{-3}$, where H$_2$ cooling is no longer efficient. However, \citet{fernandez14} demonstrate that the cold flows dissipate their energy at too low densities for the gas to enter the `zone of no return', where H$_2$ cooling becomes unimportant. Similarly, \citet{tl14} suggest that streaming velocities may suppress halo collapse and molecule formation until the atomic cooling threshold is surpassed. However, \citet{vhb14a} demonstrated that even in this case the density does not become high enough to suppress H$_2$ cooling. \citet{johnson14a} found that the ionizing flux that likely accompanies the LW radiation may boost the electron fraction and enhance molecule formation, which increases the critical LW flux. However, they noted that this should only occur in rare cases, since the IGM is more transparent to LW radiation, and the halo is likely able to shield itself from ionizing radiation. In a similar mechanism, X-rays may increase the critical LW flux by increasing the electron fraction and facilitating the formation of H$_2$ \citep{it14}.

If H$_2$ cooling is indeed suppressed until the halo reaches a virial temperature of $\simeq 10^4\,{\rm K}$, the gas contracts nearly isothermally and becomes gravitationally unstable at a Jeans mass of $\simeq 10^5-10^6\,{\rm M}_\odot$. The angular momentum of the contracting cloud is redistributed by turbulence and bar-like instabilities, such that the collapse proceeds nearly unhindered until the cloud becomes optically thick to continuum emission and a protostar forms \citep{oh02, kbd04, bvr06, ln06, wta08, bs09, csb13, latif13a, pjh13}. Due to the high temperature of the gas, the time-averaged accretion rate onto the protostar is of order $1\,{\rm M}_\odot\,{\rm yr}^{-1}$, eventually resulting in the formation of a super-massive star. Once enough mass has been accreted, a general relativistic instability develops and a fraction of the star collapses into a BH (see Section~\ref{sec_sn}). As opposed to minihalos, photoheating is not able to significantly suppress accretion, since the virial temperature of the halo is of order the temperature to which the gas is heated, and momentum transfer by photons only mildly reduces the collapse rate \citep{johnson11, johnson12}.

The evolution of the central cloud beyond the initial collapse of the gas was investigated in a number of studies. \citet{rh09} used adaptive mesh refinement simulations with a maximum resolution of $\simeq 0.01\,{\rm pc}$, $16$ cells per Jeans length, and employed a pressure floor to avoid artificial fragmentation. Due to the limited resolution, the collapse stalled at a density of $n_{\rm H}\simeq 10^9\,{\rm cm}^{-3}$, followed by the formation of a massive disk around the central hydrostatic core. In one of the three halos investigated, the cloud fragmented into three distinct clumps. \citet{latif13e} employed sink particles above a density of $n_{\rm H}\simeq 10^6\,{\rm cm}^{-3}$ and found that some of the clouds were prone to fragmentation. In a follow-up study, \citet{latif13d} investigated nine different halos using $64$ cells per Jeans length, and employed a pressure floor at a similar density as \citet{rh09}. They found that even though fragmentation occurs, the central object continues to grow via turbulent accretion and mergers at a rate of $\simeq 1\,{\rm M}_\odot\,{\rm yr}^{-1}$. At this rate, a super-massive star with $\simeq 10^6\,{\rm M}_\odot$ would form after only $\simeq 1\,{\rm Myr}$ \citep[see also][]{ih14}. Similar fragmentation and accretion was found in the high-resolution simulations of \citet{rjh14}. In a complementary approach, \citet{iot14} included the most comprehensive chemical and thermal model to date. Although they do not start from cosmological initial conditions and terminate the simulation once the density reaches $n_{\rm H}\simeq 10^{17}\,{\rm cm}^{-3}$, they find a minimum Jeans mass of $\simeq 0.2\,{\rm M}_\odot$, which is more than an order of magnitude higher than in other star formation environments.

\begin{figure}
\centering
\includegraphics[width=13cm]{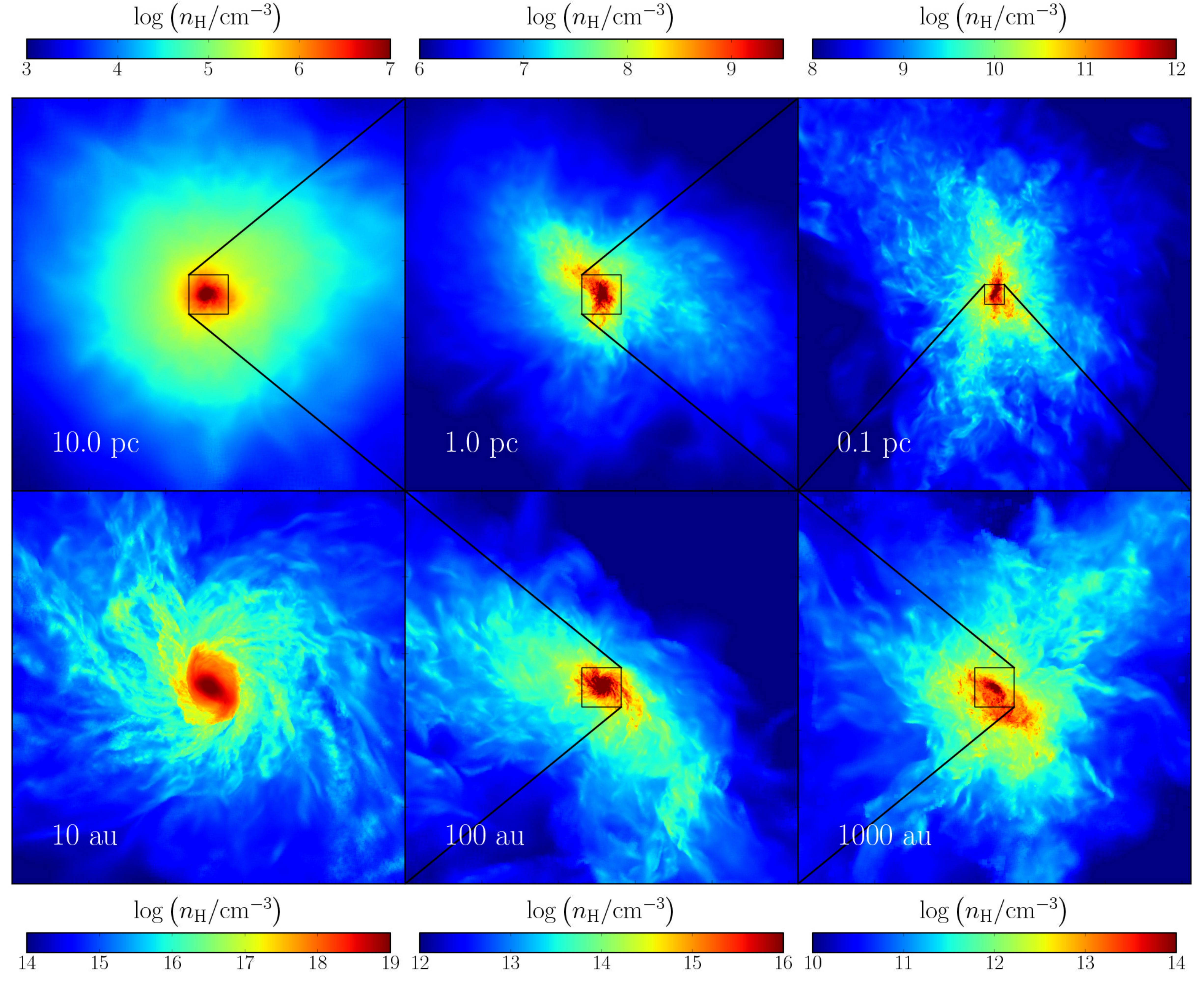}
\caption{Zoom-in on the gas cloud that forms at the center of an atomic cooling halo, showing the number density of hydrogen nuclei. Clockwise from the top left, the width of the individual cubes are $10\,$pc, $1\,$pc, $0.1\,$pc, $1000\,$au, $100\,$au, and $10\,$au. The cloud has an irregular morphology that continues to change shape and orientation throughout the collapse. The filamentary structure indicates that turbulence is present on all scales. Adapted from \citet{becerra14}.}
\label{fig_ah}
\end{figure}

One of the highest resolution simulations of the collapse of gas in atomic cooling halo was presented by \citet{becerra14}. This simulation employed a somewhat simpler chemical model than \citet{iot14}, but followed the evolution of the central gas cloud well beyond its initial collapse (see Figure~\ref{fig_ah}). In analogy to previous studies, the disk fragmented into a protostellar system with $5-10$ members, with the central protostar accreting at a rate of $\simeq 1\,{\rm M}_\odot\,{\rm yr}^{-1}$. Due to the high computational cost of the simulation, the system could only be evolved for $18\,{\rm yr}$. Nevertheless, the central protostar had already grown to $\simeq 15\,{\rm M}_\odot$. Near the end of the simulation, a second clump collapsed at about $150\,{\rm au}$ from the primary clump, potentially resulting in the formation of a wide binary system. However, the accretion rate onto the central  protostars in both clumps remains extremely high, showing that fragmentation does not prevent the rapid growth of the central objects.

\section{Empirical signatures} \label{sec_es}

A number of studies have predicted that the first stars and galaxies have distinct observational signatures. One of the most promising signatures stems from Population~III SNe \citep[e.g.][]{desouza14a}. The resulting lightcurves may show the characteristics of core-collapse SNe \citep{mw13, whalen13b, whalen13c}, pair-instability SNe \citep{scannapieco05c, wl05, hummel12, pkl12, desouza13b, whalen13a, whalen13f, whalen14}, gamma-ray bursts \citep{bl02b, cs02, its04, natarajan05, bl06, hirose06, belczynski07, ioc07, nb07, salvaterra07, sc07, salvaterra08, kb10, mr10, campisi11, dyi11, tsm11, desouza12, nakauchi12, wang12, mcz13, elliott14, mb14a, mesler14}, or even super-massive stars \citep{whalen13e, whalen13d, whalen13g}. More exotic signatures include the neutrino emission accompanying a SN explosion \citep{fwh01, iocco05, nsy06, iocco08a, suwa09}, and gravitational waves from BH binaries \citep{fwh01, bbr04, kulczycki06, suwa07b, kinugawa14}.

The properties of the first stars may also be probed by their nucleosynthetic signature, which is likely imprinted in second-generation stars that form from metal-enriched gas. If they survive to the present day, their surface abundances may reflect the metal enrichment pattern of the cloud out of which they formed. This avenue of probing the first stars has been termed `stellar archeology', or `near-field cosmology' \citep{bc05, scg08, frebel10, kbb13}. A number of stars with extremely low metallicities have been found in the Galactic halo, and their abundance patterns may reflect certain types of Population~III SNe \citep{tsy99, kg01, daigne04, frebel05, karlsson05, kg05a, daigne06, karlsson06, venkatesan06, fjb07, ssf07, tumlinson07a, tumlinson07b, kjb08, fjb09, kgc12, cm14, hartwig14b, marassi14}. Most second-generation stars are in fact expected to reside within the bulge of the Galaxy instead of the halo \citep{dmm05, tumlinson06, tss08, gao10, tumlinson10}. There may also be a connection between the first stars and globular clusters \citep{pjj97, bc02, beasley03, west04, kg05b, bekki06, moore06, bekki07, boley09}, as well as extremely metal-poor stellar populations found in local dwarf galaxies \citep{rg05, gk06, moore06, rpv06, rgs08, sfs08, br09, munoz09, ricotti09, sf09, fks10, br11a, br11b, fb12, karlsson12, mb14b}. There is even a possibility of finding true Population~III stars \citep{greif11a, sb14}, even though it may be impossible to distinguish them from other metal-poor stars due to self-enrichment or mass transfer in a binary system \citep{schlattl01, fii00, weiss00, schlattl02, picardi04, suda04, weiss04, lucatello05, lst07, tumlinson07b, lst09, sf10, starkenburg14, johnson14b}. Low-mass Population~III stars may also accrete metal-enriched gas from the IGM as they move through the Galaxy \citep{st03, sty03, fjb09, jk11}.

The radiation emitted from individual Population~III stars is too faint to be detected directly \citep[but see][]{zackrisson10a, zackrisson10b}. However, the stellar emission from the first galaxies may be detected by existing and upcoming telescopes such as the Hubble Space Telescope (HST), the JWST \citep{gardner06, gardner09}, the Atacama Large Millimeter Array (ALMA), the Square Kilometer Array (SKA), and the planned extremely large $30-40\,{\rm m}$ class telescopes. A number of studies have attempted to predict the characteristic signature of the first galaxies \citep{greif09b, johnson09, johnson10, wc09, rsf10, pmb11, sfd11, zackrisson11a, zackrisson11b, wise12a, pmb13, zij13, wise14}, with a particular emphasis on the strong He\,{\sc ii} recombination lines characteristic of metal-free stellar populations \citep{ts00, ohr01, tgs01, kudritzki02, schaerer02, schaerer03, tsv03, vts03}. A significant boost to the received luminosity may be provided by gravitational lensing \citep{maizy10, wyithe11, zackrisson12, ml13, pl13, rydberg13, whalen13h}. Finally, the first galaxies may also have contributed to the near-infrared background \citep{sbk02, msf03, sf03, cooray04, cy04, kashlinsky04, dak05, kashlinsky05b, kashlinsky05c, fk06, salvaterra06, sf06, cooray12, fz13, yue13a, yue13b}.

A number of reviews summarize the observational signatures of the first stars and galaxies. The connection between Population~III stars and extremely metal-poor halo stars is reviewed by \citet{bc05}, \citet{scg08}, \citet{frebel10}, and \citet{kbb13}, while the properties of dwarf galaxies in the Local Group are reviewed by \citet{tht09} and \citet{ricotti10}. Various reviews also focus on observations of high-redshift galaxies \citep{bland-hawthorn06, bi06, se06, ellis08, robertson10}, while the near-infrared background is summarized in \citet{hd01}, \citet{kashlinsky05c}, and \citet{kashlinsky09}. The process of reionization is described in detail in \citet{lb01}, \citet{fck06}, and \citet{stiavelli09}, and the connection between primordial star formation and the 21-cm signal is discussed in \citet{fob06} and \citet{mw10}. Finally, the properties and observational signature of BH remnants from the first stars are discussed in \citet{haiman06, haiman09}, \citet{greene12}, \citet{volonteri12}, and \citet{vb12}.

\section{Summary} \label{sec_con}

The numerical frontier of the high-redshift Universe has advanced considerably over the last $10-15$ years. The turn of the millennium saw the first three-dimensional simulations of primordial star formation that started from cosmological initial conditions and included primordial chemistry networks \citep[e.g.][]{abn02, bcl02}. They established the `standard model' of primordial star formation, in which the physics of H$_2$ cooling leads to a characteristic stellar mass of $\sim 100\,{\rm M}_\odot$. However, increasingly sophisticated simulations have begun to refine this picture. The inclusion of additional chemical and thermal processes as well as numerical and technological headway have allowed the entire collapse process to be modeled self-consistently \citep{yoshida06b, yoh08}. A key insight gained from these simulations is that primordial gas clouds are prone to fragmentation, but that the secondary protostars rapidly migrate to the center of the cloud and merge with the primary \citep{clark11b, greif12}. The most likely scenario therefore appears to be the formation of a massive central star or a binary system, surrounded by a number of significantly less massive stars.

Studies that investigated the influence of the radiation from the central protostar found that photoheating terminates accretion onto the central star and leads to the formation of Population~III stars with a wide range of masses. This is a result of their varied formation environments \citep[e.g.][]{hosokawa11, sgb12, hirano14}. Recent simulations have also begun to include magnetic fields, and found that they are amplified to dynamically important levels via the turbulent dynamo during the initial collapse \citep[e.g.][]{sur10, peters12, turk12}. A strong magnetic field enhances the rate of angular momentum transport and reduces the susceptibility of the gas to fragmentation \citep[e.g.][]{md13, peters14}. A number of other physical processes may play a role as well. These include HD cooling, cosmic rays, streaming velocities, DM annihilation, and alternative cosmologies.

The second step in the hierarchy of structure formation is the formation of the first galaxies in atomic cooling halos. The first high-resolution simulations focused on the virialization of the gas in the host halo \citep[e.g.][]{greif08, wa07b}. Later on, they included star formation recipes that modeled the radiative, mechanical and chemical feedback from Population~III stars in their progenitor halos \citep[e.g.][]{wa08b, greif10}. The UV radiation of massive Population~III stars dissociates molecules and photoheats the gas to $\simeq 10^4\,{\rm K}$ \citep[e.g.][]{abs06, jgb07, on08, whm10}. The metals dispersed by their SNe mix with primordial gas as they recollapse into the halo of the nascent galaxy \citep[e.g.][]{wise12a, ritter14}. This leads to the formation of the first metal-enriched stellar clusters, with an IMF that resembles that of our Galaxy \citep[e.g.][]{to06, dopcke13, smb14b}. The first galaxies started the process of reionization \citep[e.g.][]{wc07, ahn12, salvadori14}, and some may have seeded the first quasars by forming massive BHs in halos subjected to a strong LW background \citep[e.g.][]{omukai01, bl03b, dijkstra08, latif13d, iot14, rjh14}.

Thanks to advances in computer technology and simulation methods, our understanding of primordial star and galaxy formation has rapidly increased. The well-known initial conditions provided by observations of the CMB make this field particularly attractive. The underlying equations are well known, such that obtaining an accurate solution is merely a matter of complexity. Based on the current rate of progress, the first simulations of primordial star formation that include radiative transfer as well as magnetic fields will become possible within the next five years. This nicely coincides with the commissioning of the next generation of ground- and space-based telescopes, such as the upcoming 30--40\,m class telescopes or the JWST.

\section*{Competing interests}

The author declares that he has no competing interests.

\section*{Acknowledgements}

The author would like to thank Volker Bromm, Anastasia Fialkov, Simon Glover, Chalence Safranek-Shrader, Dominik Schleicher, and Naoki Yoshida for comments and suggestions that helped improve the manuscript.

\end{document}